\documentclass[aps,prb,longbibliography,twocolumn,floatfix,superscriptaddress,nofootinbib]{revtex4-2}

\usepackage{graphicx}
\usepackage{amsmath}
\usepackage{braket}
\usepackage{amssymb}
\usepackage{amsmath}
\usepackage{mathtools}
\usepackage{makeidx}
\usepackage{amsfonts}
\usepackage{bm}
\usepackage{xcolor} 
\usepackage{tikz}
\usepackage{tikz-3dplot}
\usetikzlibrary{3d,shapes.arrows,decorations.pathreplacing, arrows.meta, positioning, matrix}
\usepackage[colorlinks=true, linkcolor=blue, citecolor=blue, urlcolor=blue]{hyperref}
\usepackage[normalem]{ulem}

\definecolor{PG}{RGB}{255,10,20}
\definecolor{change}{RGB}{122,15,255}
\newcommand\Mycomb[2][^n]{\prescript{#1\mkern-0.5mu}{}C_{#2}}

\begin{document}

\title{Phase Transitions in a $p$ spin Glass Model: A Numerical Study}
\author{Prerak Gupta}
\email{prerak23@iiserb.ac.in}
\affiliation{Department of Physics, Indian Institute of Science Education and Research, Bhopal, Madhya Pradesh 462066, India}
\author{Auditya Sharma}
\email{auditya@iiserb.ac.in}
\affiliation{Department of Physics, Indian Institute of Science Education and Research, Bhopal, Madhya Pradesh 462066, India}

\author{Bharadwaj Vedula}
\affiliation{Department of Physics, Indian Institute of Science Education and Research, Bhopal, Madhya Pradesh 462066, India}
\author{J.Yeo}
\affiliation{Department of Physics, Konkuk University, Seoul 05029, Korea}
\author{M.A. Moore}
\affiliation{Department of Physics and Astronomy, University of Manchester, Manchester M13 9PL, United Kingdom}

\date{\today}

\begin{abstract}
We investigate the balanced $M=4$, $p=4$ spin-glass model for a one-dimensional long-range proxy for the finite dimensional short-range $p$-spin glass model to examine the nature of the glass transition beyond mean-field theory. We perform large-scale Monte Carlo equilibrated simulations for both fully connected and power-law diluted versions of the model. The critical temperatures extracted from the finite-size scaling (FSS) analysis of spin-glass susceptibility are in good agreement with theoretical predictions for $\sigma = 0, 0.25$, and 0.55. For these values of the long-range exponent $\sigma$ (which is the power of the decrease of the interactions between the spins with their separation), one might have expected that mean-field theory would provide a good description of the system.  However, the spin-overlap distribution and the value of the $\lambda$-parameter do not provide numerical evidence for a one-step replica symmetry breaking (1RSB) phase transition. Instead, our results indicate a direct transition from the paramagnetic state to a full replica symmetry broken phase, with a renormalized value of $\lambda\equiv \omega_2/\omega_1 < 1$ suggesting a continuous FRSB transition, despite this ratio being equal to 2 at mean-field level. A value of $\lambda > 1$ is required for the discontinuous 1RSB transition. We argue that strong finite-size effects and closely spaced transition temperatures remove the expected 1RSB transition for the system sizes which we can study. For values of the exponent $\sigma = 0.85$  which roughly corresponds to a three dimensional system, we find that the renormalized value of $\lambda$  is again less than 1, with no signs of either the 1RSB transition or the continuous FRSB transition, suggesting that the Kauzmann temperature $T_K$ in three dimensions might be zero, and the complete absence of phase transitions in structural glasses. 
\end{abstract}

\maketitle

\section{Introduction}
\label{sec:introduction}

Understanding the nature of the structural glass transition remains one of the central problems in statistical physics. 
Despite decades of experimental, numerical, and theoretical investigations~\citep{Mooredro,mydosh,parisi_1983,parisi_1979,sk_model,EA_1975,Fan2023,optimisation,takahasi,metastablemin2016}, the connection between slow glassy dynamics and the existence of an underlying thermodynamic phase transition is still under active debate. 
Spin-glass models have played a central role in this discussion, as they provide well-defined theoretical frameworks in which both static and dynamical aspects of glassy behavior can be systematically explored.
A major conceptual advance came with the solution of mean-field spin-glass models through replica symmetry breaking (RSB), introduced by Parisi and collaborators~\cite{parisi_1979,parisi_1983}.
Mean-field spin-glass models with multi-spin interactions, such as the $p$-spin model, are characterized by a rugged free-energy landscape containing an exponential number of metastable states. 
Their dynamics at temperatures above the mode-coupling temperature $T_{MC}$ is described by the same equations as those of mode-coupling theory (MCT)~\cite{Goetze1984,Gotze_1992,BOUCHAUD1996243}. 
For $p \ge 3$, these models exhibit a discontinuous glass transition associated with a one-step replica symmetry breaking (1RSB) solution. 
Within this framework two characteristic temperatures arise. The first is the dynamical transition temperature $T_d$ (equivalent to $T_{MC}$), where the dynamics becomes nonergodic within mean-field theory. The second is the lower thermodynamic transition temperature, the Kauzmann temperature $T_K$~\cite{Kauzmann1948TheNO}, at which the configurational entropy vanishes.

A closely related phenomenology appears in the mean-field Potts glass~\cite{Claudio_Brangian_2002}. For sufficiently large number of Potts states $p_{\text{potts}}$, the model exhibits a discontinuous glass transition with the same qualitative features as the $p$-spin model, including a 1RSB solution and a separation between $T_d$ and $T_K$. 
These results provided one of the earliest concrete realizations of what later became known as the Random First-Order Transition (RFOT) scenario~\cite{PhysRevA.40.1045}. 
Within RFOT, the glass transition of structural glasses is interpreted as the finite-dimensional counterpart of the mean-field transition found in $p$-spin and Potts models, with finite-range effects primarily modifying the dynamics while preserving the underlying thermodynamic structure. Beyond the mean-field limit, however, the dynamical transition at $T_d$ is expected to disappear due to activated processes over finite barriers separating metastatble states. 
A great deal of effort has been made to understand the existence and nature of the thermodynamic glass transition at $T_K$ in finite dimensions. A recent numerical simulation ~\cite{berthier:18b, berthier:2025} suggests that $T_K=0$ in two dimensions.

The mean-field description of $p$-spin glasses was extended by Gardner~\cite{GARDNER1985747} and Gross et al.~\cite{gross:85} respectively, where a second transition was found to occur within the 1RSB glass phase at sufficiently low temperatures. These systems thus
display two distinct transitions. At a temperature $T_1$, there is a discontinuous transition with a jump in the order parameter but no latent heat, corresponding to the emergence of a one-step replica symmetry breaking (1RSB) phase. At a lower temperature $T_2$, this 1RSB phase becomes unstable and a continuous transition to a full replica symmetry breaking (FRSB) phase occurs. This second instability is the Gardner transition, where metastable states fragment into a 
hierarchical organization as has 
been discussed in Ref.~\cite{Berthier_Biroli_Charbonneau_Corwin_Franz_Zamponi_2019}.

In recent years, considerable attention has been devoted to long-range interacting models, beginning with the model introduced by Kotliar \textit{et al.}~\cite{PhysRevB.27.602}. 
A key advantage of these models is that varying the power-law exponent $\sigma$ effectively mimics changing the spatial dimension $d$ in short-range systems. 
The quantitative relation between these parameters has been studied in detail~\cite{larson,PhysRevLett.103.267201,long_short_connection,youngmori}.
Although several works have investigated $p$-spin glasses in finite dimensions~\cite{PhysRevB.58.12081,PhysRevB.54.9756,PhysRevLett.81.1698,PhysRevB.59.1036,Mooredro,PhysRevLett.96.095701,PhysRevE.60.58,PhysRevLett.96.137202}, clear numerical evidence for a stable 1RSB phase in such settings is lacking. 
Similarly, numerical studies of finite-dimensional disordered Potts models with $p_{\text{potts}}$ up to 10 have not found convincing evidence for a discontinuous glass transition~\cite{Brangian_2003,Brangian_2002,PhysRevB.74.104416,Alvarez_2010}.
On the other hand, the infinite-range $p=3$ model shows evidence consistent with a 1RSB scenario~\cite{Billoire_2005}.

Implementing pure $p$-spin interactions on finite dimensional lattices is technically challenging~\cite{RIEGER1992279,PhysRevB.54.9756,PhysRevLett.81.1698,PhysRevB.59.1036}, since genuine multi-spin couplings do not arise naturally from simple short-range Hamiltonians and are difficult to realize in a controlled way. In contrast, the $M – p$ construction provides an effective lattice realization of $p$-spin–like physics while retaining analytical tractability. It embeds multi-spin interactions through multiple spin components per site, allowing the model to be defined naturally on a lattice with short-range couplings. As a result, $M$–$p$ models offer a practical framework for investigating physics beyond the fully connected limit~\cite{larson,PhysRevB.85.100405,caltagirone,PhysRevE.86.052501}.
In Ref.~\cite{jyeo}, two of us have studied the $p=4$ model using the Landau expansion to quintic order. At mean field level, the model shows a continuous transition to FRSB phase for small values of $M<M_{**}(<3)$. For larger values of $M$, the model undergoes a continuous ($M_{**}<M<3$) or discontinuous ($M>3$) transition to 1RSB followed by the Gardner transition to FRSB
as the temperature is reduced. Interestingly, the FRSB form when there is a continuous phase transition from the paramagnetic state is different from 
that found in Refs.~\cite{gross:85,PhysRevE.88.032135} for the state reached from the discontinuous 1RSB state.
Motivated by these results, we have studied by simulations the $p=4$ spin-glass model with $M=4$~\cite{caltagirone,jyeo} as a one-dimensional proxy in the non-extensive regime ($0 \leq \sigma < 1/2$) on a fully connected graph for $\sigma = 0$ and $0.25$, where $\sigma = 0$ corresponds to the SK limit~\cite{sk_model} of this model. 
We also present a systematic investigation of the power-law diluted interaction model with fixed average coordination number $z$~\cite{larson,PRl_2009_AH}, covering both the mean-field and non-extensive regimes ($0 \leq \sigma \leq 2/3$). 
For the diluted case, $\sigma = 0$ reduces to the Viana-Bray limit~\cite{Viana_Bray_1985}.
Through this work, we address several open questions concerning the nature of the phase transition in these models. 
Our results strongly support the claim~\cite{Mori_2011} that, in the non-extensive regime, the long-range spin glass model is equivalent to its SK limit ($\sigma = 0$) when the coupling variance is scaled to remain constant with system size and $\sigma$, at least as regards the continuous transition from the paramagnetic state to the FRSB state.  This scenario of Mori was later confirmed numerically by Wittmann and Young~\cite{youngmori} for both fully connected and diluted versions of the 1-D long-range Ising spin glass.

We perform large-scale Monte Carlo simulations to determine the transition temperature in both the fully connected and the power-law diluted cases with $z=6$. 
Our numerical estimates of the transition temperatures agree within statistical error with those reported by Yeo and Moore~\cite{jyeo}. 
In addition, we present analytical results for the Viana-Bray limit, which are also in good agreement with numerical findings. 
To probe the possible 1RSB behavior, we analyze the spin-overlap distribution function. 
However, we do not observe clear signatures of the expected one-step RSB. 
To further clarify the nature of the transition, we examine the $\lambda$-parameter, the ratio of the cubic coefficients in the static replicated Gibbs free energy Landau functional ($\lambda=\omega_2/\omega_1$)~\cite{Parisi_ratio_2013}. 
This quantity provides insight into the character of the phase transition near criticality. A discontinuous transition is expected if the value of $\lambda > 1$. For the $p=4$, $M= 4$ model which we study in this paper, $\lambda = 2$ at mean-field level, so a 1RSB discontinuous transition would be expected.
Our results suggest that strong finite-size effects, arising from the  limitations in accessible system sizes,  remove the expected 1RSB features. They also drive the renormalized values of $\lambda$ to below $1$, which is consistent with the apparent absence of the 1RSB transition.
In the diluted model within the mean-field regime, fluctuation effects and/or finite size effects also appear to drive the renormalized value $\lambda$ below unity. The only transition would seem to be a continuous transition as found at mean-field level when $M < M^{**}$ parameter ~\cite{jyeo,Franz1999}. This transition is in the universality class of the Ising model de Almeida-Thouless transition~\cite{Bharadwaj:2025,BharadwajE:2025}.  We would not therefore expect to see the continuous transition to a state with FRSB below six dimensions (or in the one-dimensional proxy model for $\sigma > 2/3$). We have studied the diluted model at a value of $\sigma = 0.85$ (which corresponds in the short-range model to a dimensionality $d$ of around $3$) and we see neither the continuous transition  to FRSB nor the 1RSB transition. The  renormalized value of $\lambda $ is again less than  $1$ which is consistent with the absence of a 1RSB transition. Thus we would expect that in three dimensions the Kauzmann temperature $T_K$ to also be zero, just as has been argued is the case in two dimensions~\cite{berthier:18b,berthier:2025}.

Overall, our findings favor a direct transition from the replica-symmetric paramagnetic phase to a full replica symmetry broken (FRSB) phase, with no numerical evidence for an intermediate 1RSB transition. The behavior of the spin-overlap distribution and the associated cumulants are all consistent with this interpretation for the interaction types studied. 
We interpret these results as being consistent with the strong finite-size effects that may merge nearby transition temperatures and mask the RS-1RSB separation in accessible system sizes. 
For the system sizes accessible in our simulations, these effects modify the effective values of the coefficients in the replicated Gibbs free-energy expansion, $\omega_1$ and $\omega_2$. In particular, they lead to an effective ratio ($\lambda< 1$), which suppresses the 1RSB transition. We expect that if simulations could be performed for significantly larger system sizes in the fully connected model, the 1RSB transition would eventually emerge. Although the numerical simulations include the effects of higher-order terms in the expansion, the primary consequence of finite system size is to shift the effective values of these coefficients away from their large system size limits, i.e., the values used in our earlier mean-field study~\cite{jyeo}. These corrections arise from finite size contributions beyond the steepest-descent approximation.
For the diluted model with nonzero ($\sigma$), we also do not observe clear evidence of a 1RSB transition in our simulations. However, it remains unclear whether this absence is due to finite-size effects or to fluctuations that become important away from the mean-field limit. Consequently, based on the present numerical results, we cannot determine whether the 1RSB transition would reappear in the thermodynamic limit.

The paper is organized as follows. 
In Sec.~(\ref{sec:Model}), we introduce the one-dimensional four-leg ladder model for both the fully connected and power-law diluted cases in the non-extensive regime. 
Section~(\ref{sec:Method}) describes the equilibration criteria and the observables analyzed in the simulations. 
In Sec.~(\ref{sec:Results}), we present a detailed discussion of our numerical and analytical results. 
Finally, Sec.~(\ref{sec:summary and conclusion}) summarizes the main conclusions.

\section{Models}
\label{sec:Model}
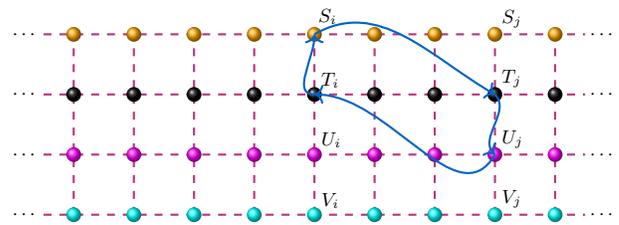
\begin{figure}
\begin{tikzpicture}[scale=0.8, every node/.style={transform shape}]
            \definecolor{bgcolor}{RGB}{185,56,124}
            \definecolor{level4}{RGB}{255,165,0}
            \definecolor{cyclecolor}{RGB}{0,100,200}
            
            \foreach \y in {0,1,2,3} {
                \draw[bgcolor, dashed, thick] (-0.5,\y) -- (8.5,\y);
                \node at (-0.8,\y) {$\cdots$};
                \node at (8.8,\y) {$\cdots$};
            }
            
            \foreach \x in {0,1,...,8} {
                \draw[bgcolor, dashed, thick] (\x,0) -- (\x,3);
                \shade[ball color=cyan] (\x,0) circle(3.5pt);
                \shade[ball color=magenta] (\x,1) circle(3.5pt);
                \shade[ball color=black] (\x,2) circle(3.5pt);
                \shade[ball color=level4] (\x,3) circle(3.5pt);
            }
            
            \node[above right] at (3.95,3.05) {$S_i$};
            \node[above right] at (4,2) {$T_i$};
            \node[above right] at (4,1) {$U_i$};
            \node[above right] at (4,0) {$V_i$};
            \node[above right] at (7,3) {$S_j$};
            \node[above right] at (7,2) {$T_j$};
            \node[above right] at (7,1) {$U_j$};
            \node[above right] at (7,0) {$V_j$};
            
            \draw[cyclecolor, ->, thick, out=30, in=150] (4,3) to (7,2);
            \draw[cyclecolor, ->, thick, out=-45, in=135] (7,2) to (7,1);
            \draw[cyclecolor, ->, thick, out=-120, in=0] (7,1) to (4,2);
            \draw[cyclecolor, ->, thick, out=180, in=-90] (4,2) to (4,3);
            
        \end{tikzpicture}%
        \caption{The lattice (dashed lines) consists of a four-leg
          ladder with an Ising spins $S_i$, $T_i$, $U_i$, $V_i$ on the
          $i^{th}$ rung. An interaction couples any two spins
          sitting on the $i^{th}$ rung with any two of the spins from the
          $j^{th}$ one. The solid line (in blue) shows the interaction
          involving $S_i$, $T_i$, $T_j$, and $U_j$.}
\label{fig:schematic}
\end{figure}

In this paper, we investigate the balanced $M-p$ model with $M = 4$ and $p = 4$. 
The system consists of a four-leg ladder with $L$ rungs, where each rung hosts four Ising spins, denoted by $S$, $T$, $U$, and $V$ (see Fig.~\ref{fig:schematic}). 
Since the interactions involve four spins, the model corresponds to $p=4$. 
According to the interaction scheme, any pair of spins on one rung can interact with any pair of spins on another rung. 
Consequently, for every interacting pair of rungs, the Hamiltonian contains a total of $(\Mycomb[4]{2})^2$ distinct four-spin interaction terms (see Eq.~\eqref{eq:Hamiltonian_M4p4_pair} for details). 
The Hamiltonian can be written in the compact form

\begin{equation}
\label{Hamil_compact}
\mathcal{H} = - \sum_{\langle i,j \rangle} \sum_{(a<b)} \sum_{(c<d)} 
J_{ij}^{(ab,cd)} \, S_{i}^{a} S_{i}^{b} S_{j}^{c} S_{j}^{d},
\end{equation}
where $i,j = 1,2,\dots,L$ label the interacting pair of rungs. 
The spin indices satisfy $(a,b,c,d) \in \{1,2,3,4\}$ with $a<b$ and $c<d$. 
The four Ising spins on each rung are identified as shown in Fig.~\ref{fig:schematic}:
\[
S_{i}^{1} = S_{i}, \qquad 
S_{i}^{2} = T_{i}, \qquad 
S_{i}^{3} = U_{i}, \qquad 
S_{i}^{4} = V_{i}.
\]
The coupling constants $J_{ij}^{(ab,cd)}$ are quenched, independent random variables drawn from a Gaussian distribution with zero mean and variance which is 
subjected to the choice of model, as mentioned in the following subsections.

\subsection{Power-Law Fully Connected Interaction}
In the fully connected version of the model, each rung interacts with every other rung in the system. The strength of these interactions decreases with the distance between the rungs and is controlled by the parameter $\sigma$, which appears in the random couplings. The case $\sigma = 0$ corresponds to the Sherrington Kirkpatrick (SK) limit.
The couplings $J_{ij}$ are defined as
\begin{equation}
J_{ij} = C(\sigma)\,\frac{\epsilon_{ij}}{r_{ij}^{\sigma}},
\label{eq:coupling}
\end{equation}
where the variables $\epsilon_{ij}$ are independent Gaussian random
numbers with zero mean and unit variance. The distance between rungs
$i$ and $j$, arranged on a ring, is
\begin{equation}
    r_{ij} = \left(\frac{L}{\pi}\right)\sin\!\left(\frac{\pi|i-j|}{L}\right).
\end{equation}
The normalization constant $C(\sigma)$ is chosen such that the variance of the couplings obeys
\begin{equation}
\sum_{j \neq i} \big[J_{ij}^{2}\big]_{\mathrm{av}}
= 1/(M^{p-1})
= C(\sigma)^{2}\sum_{j \neq i}\frac{1}{r_{ij}^{2\sigma}},
\label{eq:c_sigma}
\end{equation}
where $[\cdots]_{\mathrm{av}}$ denotes an average over disorder realizations. 
To ensure this normalization, one can draw the $\epsilon_{ij}$ from a
normal distribution $\mathcal{N}(0,1)$, and fix the constant $C(\sigma)$ such that the condition in Eqn.~\eqref{eq:c_sigma} holds. Eq.~\eqref{eq:c_sigma} therefore provides a direct way to compute $C(\sigma)$ for any given system size $L$ and interaction parameter $\sigma$. In the non-extensive regime, this constant decreases with increasing $L$ and vanishes in the limit $L \to \infty$. Table \ref{tab:table_fully_conn} lists the simulation parameters used in the fully connected model.
\begin{table}[h!]
\centering
\begin{minipage}{0.48\textwidth}
\centering
\renewcommand{\arraystretch}{1.5}
\setlength{\tabcolsep}{10pt}
\caption[Simulation parameters for different values of $\sigma$]{Simulation parameters for the fully connected model. Here $N_{\text{samp}}$ is the number of disorder samples, $N_{\text{sweep}}$ is the total number of Monte Carlo sweeps, $T_{\text{min}}$ and $T_{\text{max}}$ are the lower and upper limits of the simulated temperature range, and $N_T$ is the number of temperatures. The last column lists the value of $C(\sigma)$ obtained from Eq.~\eqref{eq:c_sigma}.}
\resizebox{\textwidth}{!}{
\begin{tabular}{cccccccc}
\hline\hline
$\sigma$ & $L$  & $N_{\text{samp}}$ & $N_{\text{sweep}}$ 
& $T_{\text{min}}$ & $T_{\text{max}}$ & $N_T$ & $C$ \\[0.2cm]
\hline
0.0  &   8   & 10000 & 16384   & 0.15 & 0.6 & 17 & 0.377964  \\[0.2cm]
0.0  &  16   & 12000 & 32768   & 0.15 & 0.6 & 17 & 0.258199  \\[0.2cm]
0.0  &  32   & 10000 & 32768   & 0.2 & 0.6 & 17 & 0.179605  \\[0.2cm]
0.0  &  64   & 9000 & 65536   &  0.2 & 0.6 & 17 & 0.125988  \\[0.2cm]
0.0  & 128   & 10000 & 524288  & 0.2 & 0.6 & 17 & 0.088736  \\[0.2cm]
0.0  & 256   & 4000 & 524288  & 0.2 & 0.6 & 17 & 0.062622  \\[0.2cm]
0.25 &   8   & 10000 & 32768   & 0.15 & 0.6 & 17 & 0.428470  \\[0.2cm]
0.25 &  16   & 17000 & 32768   & 0.15 & 0.6 & 17 & 0.334941  \\[0.2cm]
0.25 &  32   & 8000 & 65536   & 0.2 & 0.6 & 17 & 0.269035  \\[0.2cm]
0.25 &  64   & 9000 & 65536   & 0.2 & 0.6 & 17 & 0.219535  \\[0.2cm]
0.25 & 128   & 7000 & 524288  & 0.2 & 0.6 & 17 & 0.180915  \\[0.2cm]
0.25 & 256   & 3550 & 524288  & 0.2 & 0.6 & 17 & 0.150044  \\[0.2cm]
\hline\hline
\end{tabular}}
\label{tab:table_fully_conn}
\end{minipage}
\end{table}

\subsection{Power-Law Diluted Interaction}

In the diluted version of the model, each rung interacts only with a
finite number of other rungs on average, rather than with all of
them. The interactions are chosen such that the probability of forming
a bond between two rungs follows a power-law decay. To construct this
network of bonds, we proceed as follows. A rung $i$ is selected
uniformly at random, and a second rung $j$ is then chosen with
probability given by Eq.~\eqref{c_eqn}. If $i$ and $j$ are already
connected, the procedure is repeated until an unconnected pair is
found. Once an unoccupied pair $(i,j)$ is identified, a bond is
assigned between them. This process is repeated until exactly
$ N_b = L{z}/2$ bonds are created, where ${z}$ is the mean
coordination number.  The probability of forming a bond between two
rungs separated by a geometric distance $r_{ij}$ is proportional to
$r_{ij}^{-2\sigma}$.  Each interacting pair of rungs is coupled by the
independent Gaussian random couplings $J_{ij}^{}$ (with mean zero and
variance $1/M^{p-1}$), corresponding to the $p$-spin interaction.
Since the naive probability $p_{ij} \propto r_{ij}^{-2\sigma}$ can
exceed unity for small separations, we introduce a short-distance
cutoff and use the modified interaction probability
~\cite{PRl_2009_AH}
\begin{equation}\label{c_eqn}
    p_{ij} = 1 - \exp\!\left(-\frac{A}{r_{ij}^{2\sigma}}\right),
\end{equation}
where the constant $A$ is chosen such that the average number of neighbors of any rung satisfies
\begin{equation}
    z = \sum_{j=0}^{L-1} p_{ij}.
\end{equation}
In this study, the mean coordination number is fixed at $z = 6$, which uniquely determines the value of the parameter $A$. Table \ref{tab:table_powe_dil} lists the simulation parameters used in the power law diuted case.

\begin{table}[h!]
\centering
\begin{minipage}{0.48\textwidth}
\centering
\renewcommand{\arraystretch}{1.5}
\setlength{\tabcolsep}{10pt}
\caption[Simulation parameters for different values of $\sigma$]{Simulation parameters for the power-law diluted model. Here $N_{\text{samp}}$ is the number of disorder realizations, $N_{\text{sweep}}$ is the total number of Monte Carlo sweeps, $T_{\text{min}}$ and $T_{\text{max}}$ are the limits of the simulated temperature range, and $N_T$ is the number of temperatures used. The final column lists the value of $A$ in Eq.~\eqref{c_eqn}, chosen to ensure an average coordination number $z = 6$.}
\resizebox{\textwidth}{!}{
\begin{tabular}{cccccccc}
\hline\hline
$\sigma$ & $L$  & $N_{\text{samp}}$ & $N_{\text{sweep}}$ 
& $T_{\text{min}}$ & $T_{\text{max}}$ & $N_T$ & $A$ \\[0.2cm]
\hline
0.0  &   8   & 10000 & 16384    & 0.5 & 0.8 & 16 & 1.94591  \\[0.2cm]
0.0  &  16   & 10000  & 32768    & 0.5 & 0.8 & 16 & 0.51082  \\[0.2cm]
0.0  &  32   & 10000  & 65536    & 0.5 & 0.8 & 16 &  0.21511  \\[0.2cm]
0.0  &  64   & 10000  & 65536    & 0.5 & 0.8 & 16 & 0.10008  \\[0.2cm]
0.0  & 128   & 10000 & 524288   & 0.5 & 0.8 & 16 & 0.04839  \\[0.2cm]
0.0  & 256   & 6000 & 1048576  & 0.5 & 0.8 & 16 & 0.02381  \\[0.2cm]
0.0  & 512   & 6000  & 1048576  & 0.5 & 0.8 & 16 & 0.01181  \\[0.2cm]
0.0  & 1024  & 3000 & 2097152  & 0.5 & 0.8 & 16 & 0.00588  \\[0.2cm]
0.25 &   8   & 10000 & 16384    & 0.5 & 0.8 & 16 & 2.59301  \\[0.2cm]
0.25 &  16   & 10000 & 32768    & 0.5 & 0.8 & 16 & 0.87970  \\[0.2cm]
0.25 &  32   & 10000 & 65536    & 0.5 & 0.8 & 16 & 0.49110  \\[0.2cm]
0.25 &  64   & 10000 & 65536    & 0.5 & 0.8 & 16 & 0.30768  \\[0.2cm]
0.25 & 128   & 12000 & 262144   & 0.5 & 0.8 & 16 & 0.20287  \\[0.2cm]
0.25 & 256   & 8000 & 262144  & 0.5 & 0.8 & 16 & 0.16744  \\[0.2cm]
0.25 & 512   & 5000 & 1048576  & 0.5 & 0.8 & 16 & 0.09459 \\[0.2cm]
0.25 & 1024  & 2500 & 2097152  & 0.5 & 0.8 & 16 & 0.06572 \\[0.2cm]
0.55 &   8   & 10000 & 16384    & 0.5 & 0.8 & 16 & 3.85462 \\[0.2cm]
0.55 &  16   & 10000 & 32768    & 0.5 & 0.8 & 16 &  1.64275  \\[0.2cm]
0.55 &  32   & 10000 & 131072    & 0.5 & 0.8 & 16 & 1.18046  \\[0.2cm]
0.55 &  64   & 10000 & 262144    & 0.5 & 0.8 & 16 & 0.95527  \\[0.2cm]
0.55 & 128   & 10000 & 524288   & 0.5 & 0.8 & 16 & 0.81746  \\[0.2cm]
0.55 & 256   & 7000 & 524288  & 0.5 & 0.8 & 16 & 0.72316  \\[0.2cm]
0.55 & 512   & 5000 & 1048576  &0.5 & 0.8 & 16 & 0.65411 \\[0.2cm]
0.55 & 1024  & 3000 & 2097152  & 0.5 & 0.8 & 16 & 0.60129 \\[0.2cm]
0.85 &   8   & 7000 & 16384    & 0.5 & 0.8 & 16 & 6.03917 \\[0.2cm]
0.85 &  16   & 7000 & 32768    & 0.5 & 0.8 & 16 &  3.10205  \\[0.2cm]
0.85 &  32   & 7000 & 131072    & 0.5 & 0.8 & 16 & 2.65087  \\[0.2cm]
0.85 &  64   & 8000 & 262144    & 0.5 & 0.8 & 16 & 2.47900  \\[0.2cm]
0.85 & 128   & 4000 & 524288   & 0.5 & 0.8 & 16 & 2.39484  \\[0.2cm]
0.85 & 256   & 5000 & 1048576  & 0.5 & 0.8 & 16 & 2.34867  \\[0.2cm]
0.85 & 512   & 9000 & 1048576  &0.5 & 0.8 & 16 & 2.32188 \\[0.2cm]
0.85 & 1024  & 3500 & 2097152  & 0.5 & 0.8 & 16 & 2.30592 \\[0.2cm]
\hline\hline
\end{tabular}}
\label{tab:table_powe_dil}
\end{minipage}
\end{table}

\section{Method}
\label{sec:Method}
\begin{figure}[t]
    \centering
    \includegraphics[width=\columnwidth]{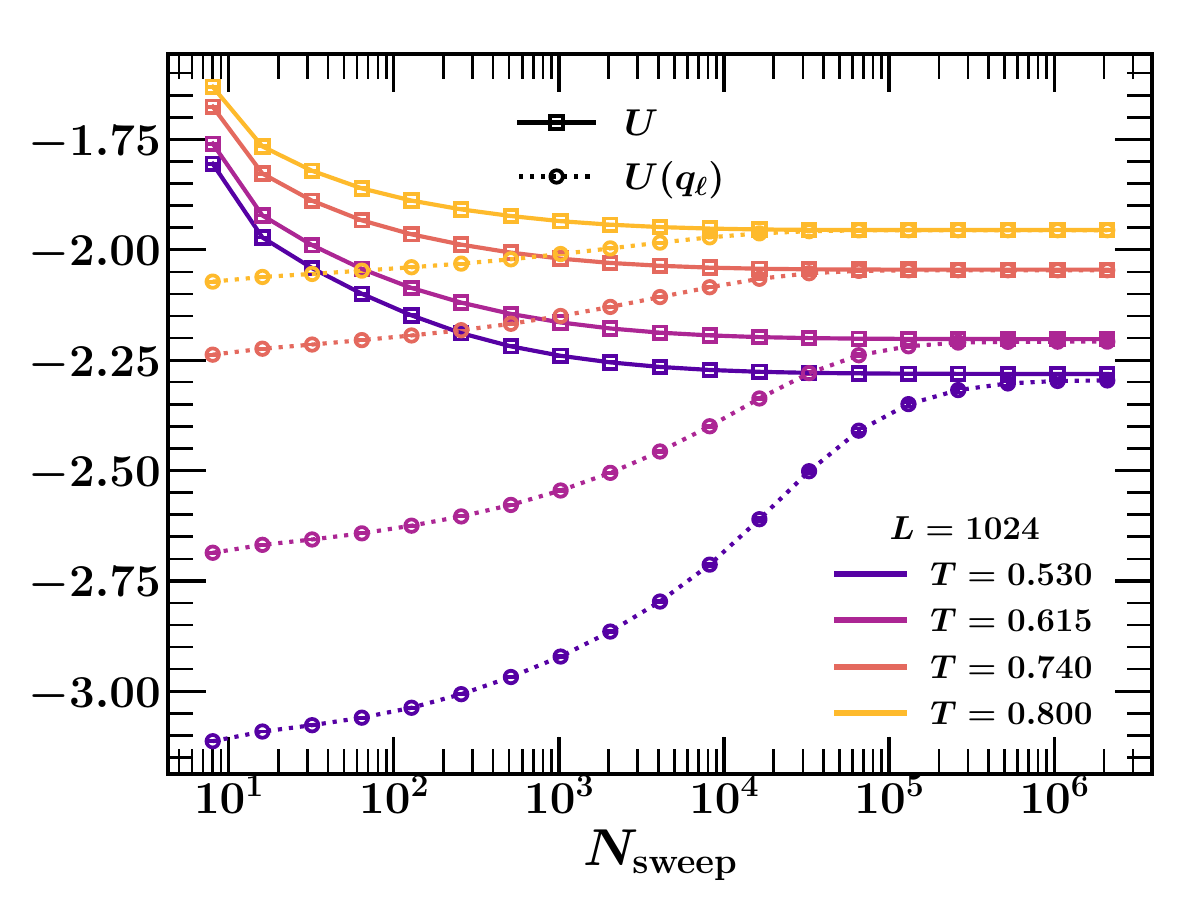}
    \caption{
    Equilibration data for system size $L=1024$ at $\sigma=0.0$ for power law diluted interaction across multiple temperatures. 
    The energy per rung ($U$) (solid lines with square markers) and $U(q_\ell)$ (see Eq.~\eqref{equilibration_equation} RHS)
    (dotted lines with circular markers) are plotted as a function of the number of Monte Carlo sweeps $N_{\mathrm{sweep}}$ 
    on a logarithmic scale. 
    The convergence of both at long Monte Carlo times indicates proper equilibration of the system. The other cases have been tested in a similar manner.
    }
    \label{fig:equilibration_L1024_sigma0}
\end{figure}
For simulations at finite temperature, we use the standard Metropolis
algorithm together with the parallel tempering (exchange Monte Carlo)
method~\cite{pixley_young,Equilibration_paper,ex_mc_Hukushima,p_mc}. Parallel
tempering improves sampling compared to single-spin flip updates, as
it allows replicas to move across temperatures and explore the rough
energy landscape more efficiently. We initialize four independent
replicas with random spin configurations (for fixed $J_{ij}$) at each
of the $N_T$ temperatures, ensuring that all starting states are
uncorrelated. After every ten Metropolis sweeps, we perform a parallel
tempering exchange step. This enables each replica to execute a random
walk in temperature space between $T_{\min}$ and $T_{\max}$, thereby
improving equilibration and sampling across the full temperature
range. 

\begin{figure}[tb]
    \centering
    \includegraphics[width=0.48\textwidth]{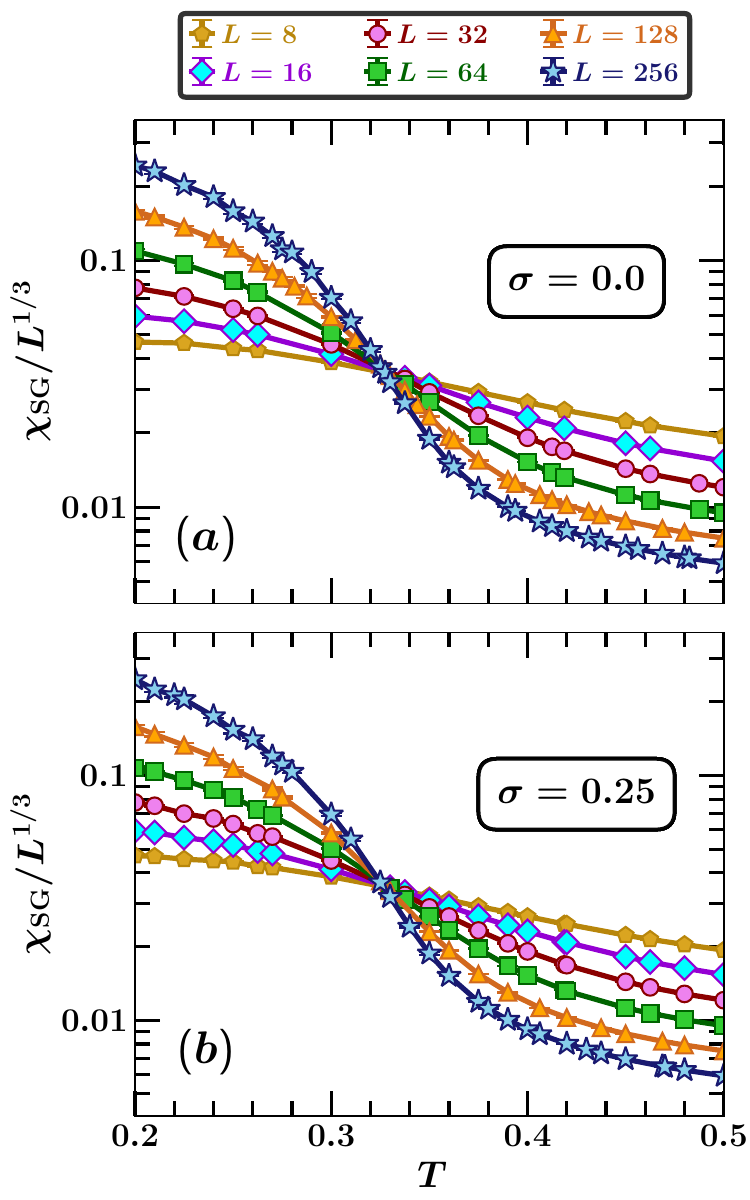}
    \caption{
        Temperature dependence of the rescaled spin glass
        susceptibility for the fully connected model.
        The plot shows data for $\sigma = 0$ and $0.25$ across multiple
        system sizes.
    }
    \label{fig:fc_collective_susceptibility}
\end{figure}
\begin{figure}[tb]
    \centering
    \includegraphics[width=0.48\textwidth]{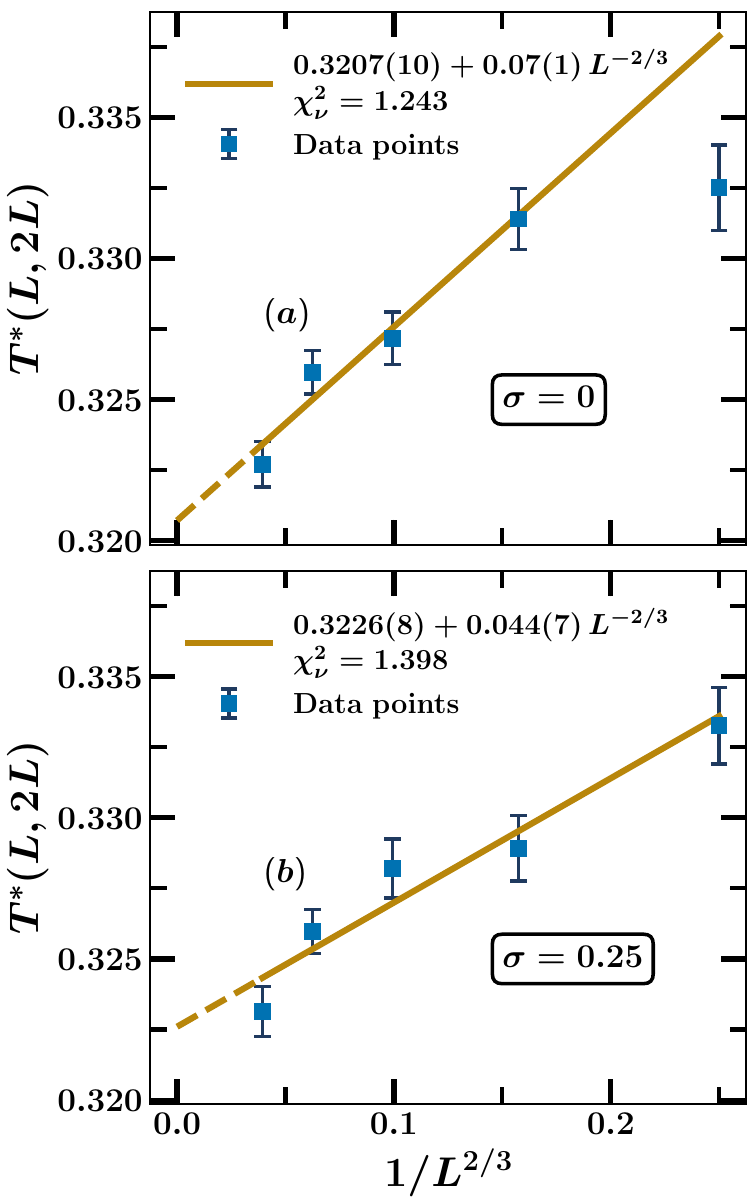}
    \caption{
        Collective finite size analysis for the fully connected model.
        The figure shows the intersection point analysis of the rescaled
        spin glass susceptibility for $\sigma = 0$ (top) and $0.25$ (bottom).
        The intersection temperatures are extrapolated as a function of the
        inverse system size to estimate the critical temperature in the
        thermodynamic limit. The reduced chi-squared $\chi_{\nu}^2$ denotes the goodness
        of the corresponding fits.
    }
    \label{fig:collective_intersection_fully_connected}
\end{figure}

\begin{figure*}[tb]
    \centering
    \includegraphics[width=\textwidth]{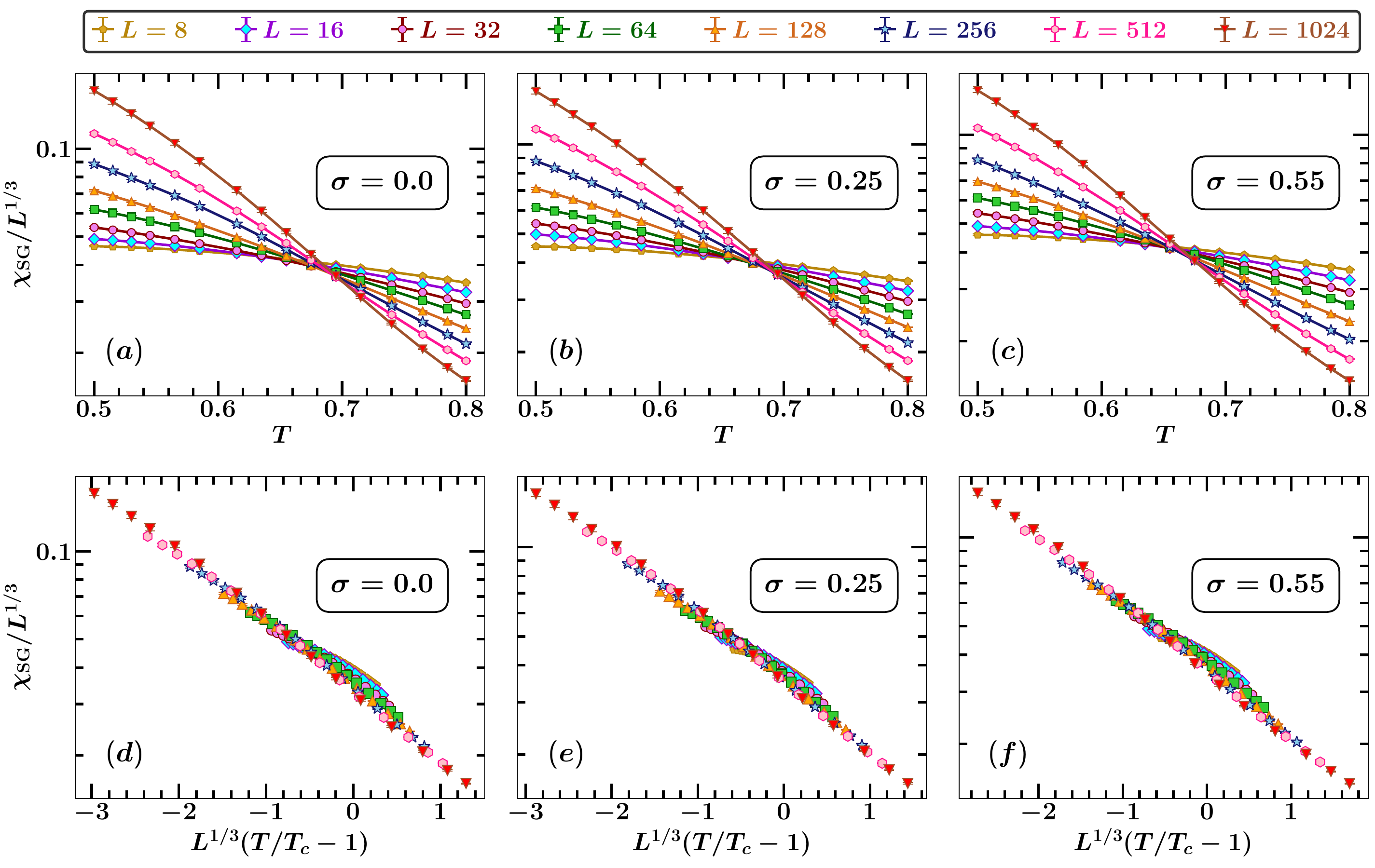}
    \caption{
    Rescaled spin glass susceptibility and finite size scaling analysis for the diluted model.
    The \emph{upper row} shows the temperature dependence of the rescaled susceptibility for
    $\sigma = 0$, $0.25$, and $0.55$ (from left to right).
    The \emph{lower row} displays the corresponding finite size scaling collapse obtained using
    the estimated critical temperatures and scaling exponents.
    The collective presentation highlights the systematic evolution of the scaling behavior
    with increasing interaction range exponent $\sigma$.
    }
    \label{fig:diluted_sus_and_collapse_collective}
\end{figure*}

\begin{figure*}
    \centering
    \includegraphics[width=\textwidth]{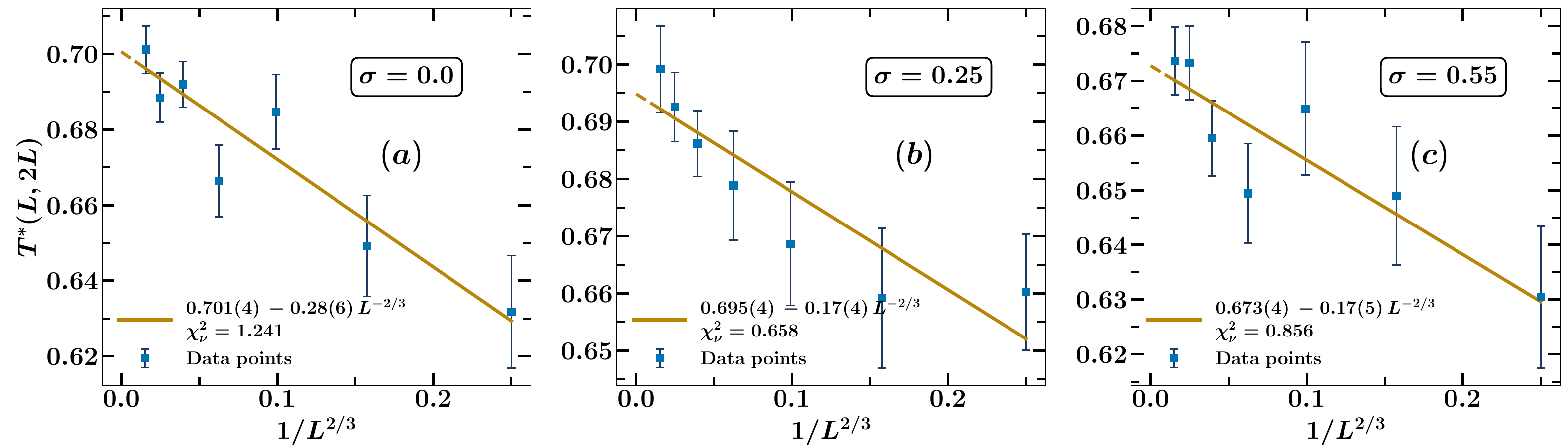}
\caption{Collective finite size analysis of the rescaled spin glass susceptibility for the power law diluted model at $\sigma = 0$, $0.25$, and $0.55$. Intersection temperatures are extrapolated versus inverse system size to obtain the thermodynamic critical temperature. Fit quality is quantified by the reduced chi-squared $\chi_{\nu}^2$.
}
    \label{fig:collective_intersection_power_law}
\end{figure*}
\begin{figure}
    \centering
    \includegraphics[width=\columnwidth]{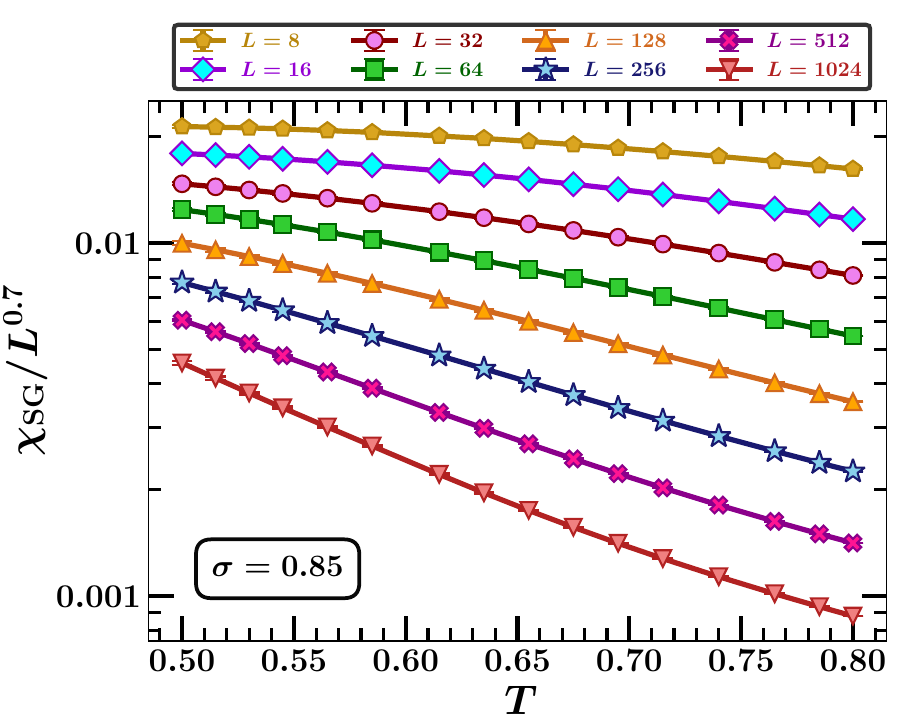} 
\caption{
Finite-size scaling behaviour of the power-law diluted model for $\sigma = 0.85$, lying in the non-mean-field regime, shown for multiple system sizes. Figure presents the temperature dependence of the rescaled spin glass susceptibility $\chi_{\mathrm{SG}}/L^{0.7}$, which shows no evidences for crossing at finite temperature.
}
    \label{fig:combined_sigma_085}
\end{figure}

To check that the system has reached equilibrium, we use a
traditional relation proposed by
Ref.~\cite{Bray_1980,Equilibration_paper} that connects the average
energy per rung to the link overlap. The equilibration condition for
the fully connected case is
\begin{equation}
U = \frac{18 \beta}{M^3}\left[ \hat{q_l}- 1 \right]_{\text{av}},
\label{eq:equi_KAS_model_with_norm}
\end{equation}
where \(U\) is the disorder-averaged energy per rung, \(q_l\) is the
link overlap, and \(N_b\) is the total number of bonds
(\(= {L(L-1)}/{2}\)). The link overlap for the fully connected case is
\begin{equation}
\hat{q_l} = \frac{\bar{z}C^2}{36N_b}\sum_{\langle ij \rangle}\frac{1}{r_{ij}^{2\sigma}}(\langle S_i T_i S_j T_j\rangle^2+ \cdots + \langle U_i V_i U_j V_j\rangle^2) 
\label{eq:link_overlap_kas}
\end{equation}
 A similar equilibration test condition for the diluted interaction case is given by
\begin{equation}\label{equilibration_equation}
    U = -\frac{36}{T} \left[ \frac{N_b}{L} \left( 1 - q_l \right) \right]_{\text{av}}, \quad \left(\text{since } \frac{\bar{z}}{2} = \frac{N_b}{L}\right),
\end{equation}
where the link overlap function for the diluted case takes the form
\begin{align}
    q_l = \frac{1}{36N_b} \sum_{\langle ij \rangle}  \left[ \langle S_i T_i S_j T_j\rangle^2  + \cdots + \langle U_i V_i U_j V_j\rangle^2 \right].
\end{align}
The equilibration test is taken to be satisfied when the two quantities on either side of the equilibration condition merge with each other within error-bars. Fig.~\ref{fig:equilibration_L1024_sigma0} shows a represenative sample of the equilibration analysis based on Eq.~\eqref{equilibration_equation} for the diluted case for the largest system size and for several temperatures used in
the simulation.
\subsection{Spin glass susceptibility and Correlation length}
The spin-glass susceptibility, which we consider as the primary observable for locating the critical point, is defined as
\begin{align}
\chi_{\mathrm{SG}}
=
\frac{1}{L M^4}
\sum_{ij}
\Bigg[
\sum_{\text{36 pairs}}
\Big(
\langle \mathcal{O}_i \mathcal{O}_j \rangle - 
\langle \mathcal{O}_i \rangle
\langle \mathcal{O}_j \rangle
\Big)^2
\Bigg]_\text{av},
\label{eq:chi_final_appendix}
\end{align}
where $\mathcal{O}_i$ denotes one of the possible pair variables
\[
\left(S_iT_i,\; S_iU_i,\; S_iV_i,\; T_iU_i,\; T_iV_i,\; U_iV_i\right).
\]
The summation over all $36$ pair combinations arises naturally from the structure of the $M=4,\;p=4$ Hamiltonian. The analytical computation of this quantity is presented in Appendix~\ref{appendix:susceptibility}. This quantity is directly related to the zero-momentum replicon propagator,
\[
\chi_{\mathrm{SG}} = G_{\mathrm{R}}(k=0),
\]
which describes the fluctuations of the replica overlap field. The replicon propagator emerges from the quadratic expansion of the replicated free energy around the replica-symmetric saddle point and represents the critical fluctuation mode of the overlap matrix. Near the transition, its inverse defines the replicon mass, and the vanishing of this mass signals the onset of spin-glass criticality~\cite{AJBray_1979}.

To investigate the critical point of the spin-glass phase transition, we therefore study the behavior of $\chi_{\mathrm{SG}}$ as a function of temperature. Although $\chi_{\mathrm{SG}}$ is not
dimensionless, its finite-size scaling (FSS) form is exactly known in
the regime ($0 < \sigma < 2/3$). In this region, one
has~\cite{PhysRevLett.103.267201,youngmori}
\begin{equation}
    \chi_{\mathrm{SG}} \sim L^{1/3}\,\tilde{\chi}\left[(T - T_c)L^{1/3}\right].
    \label{eq:chi_scaling_meanfield}
\end{equation}
In the non-mean-field regime $(\sigma > 2/3)$, the spin-glass susceptibility obeys the finite-size scaling form
\begin{equation}
    \chi_{\mathrm{SG}} \sim L^{2-\eta}\,
    \tilde{\chi}\!\left[(T-T_c)L^{1/\nu}\right],
    \label{eq:chi_scaling_nonmeanfield}
\end{equation}
where $\tilde{\chi}$ is the corresponding scaling function, $T_c$ denotes the critical temperature, and $\nu$ and $\eta$ are the standard critical exponents.
The finite-size correlation length is defined as~\cite{Equi_Ap_HGK_2005,PRl_2009_AH,PhysRevE.108.014116}
\begin{equation}
    \xi_L =
    \frac{1}{2\sin(k_{\min}/2)}
    \left[
        \frac{\chi_{\mathrm{SG}}(0)}
             {\chi_{\mathrm{SG}}(k_{\min})}
        -1
    \right]^{1/(2\sigma-1)},
\end{equation}
where $k_{\min}=2\pi/L$ is the smallest nonzero wave vector.
The corresponding finite-size scaling relation for the correlation length in the non-mean-field regime takes the form
\begin{equation}
    \frac{\xi_L}{L}
    \sim
    \mathcal{C}\!\left[
        L^{1/\nu}(T-T_c)
    \right],
    \qquad \sigma > \frac{2}{3},
\end{equation}
with $\mathcal{C}$ denoting the scaling function.
For the mean-field regime $\left(\frac{1}{2}<\sigma\leq\frac{2}{3}\right)$, the scaling form is instead given by~\cite{PRl_2009_AH}
\begin{equation}
    \frac{\xi_L}{L^{\nu/3}}
    \sim
    \mathcal{C}\!\left[
        L^{1/\nu}(T-T_c)
    \right],
    \qquad \frac{1}{2}<\sigma\leq\frac{2}{3},
\end{equation}
where the critical exponent is $\nu = \frac{1}{2\sigma-1}$.
A standard approach for estimating the transition temperature is to inspect the crossing points of the curves for $\chi_{\mathrm{SG}}/L^{1/3}$ ($\chi_{\mathrm{SG}}/L^{2-\eta}$ in non-mean field regime) for different system sizes. However, as indicated by the susceptibility plots (see Figs.~\ref{fig:diluted_sus_and_collapse_collective}), these curves do not meet at a single temperature, suggesting that corrections to the idealized scaling forms in the above equations 
plays an important role.
To understand the origin of these deviations, we recall the general FSS expression for the susceptibility near criticality~\cite{larson,aspel,PhysRevB.78.214205}:
\begin{equation}
    \chi_{\mathrm{SG}}(t, L) = L^{a} f(L^{b} t)
    + L^{-\omega} g(L^{y} t) + \cdots 
    + c_0 + c_1 t + \cdots,
    \label{eq:chi_general_fss}
\end{equation}
where $t = T - T_c$, $a = 2 - \eta$ (which equals $2\sigma - 1$), and $b = 1/\nu$.  
The term proportional to $L^{-\omega}$ captures the leading singular correction to scaling, while $c_0$ denotes the dominant analytic correction.

In the mean-field regime ($1/2 < \sigma < 2/3$), the critical exponents become independent of $\sigma$. In particular, one finds $a = 1/3$ and $b = 1/3$~\cite{PhysRevB.31.1498,refId0,PhysRevB.71.174438}. Substituting these values into Eq.~(\ref{eq:chi_general_fss}) yields
\begin{equation}
\begin{aligned}
    \chi_{\mathrm{SG}}(t, L) ={}& L^{1/3} f(L^{1/3} t)
    + L^{-\omega} g(L^{1/3} t) + \cdots \\
    &+ d_0 L^{2\sigma - 1} h(L^{1/3} t)
    + c_0 + c_1 t + \cdots,
\end{aligned}
\label{eq:chi_modified_fss}
\end{equation}
where the leading correction exponent is $\omega = 2 - 3\sigma$~\cite{PhysRevB.27.602}.  
In the non-extensive regime ($\sigma < 1/2$), the analytic correction $c_0$ becomes the dominant subleading contribution.

Taking the leading correction into account implies that the crossing temperature of the susceptibility curves for system sizes $L$ and $2L$, expressed as $T^{\ast}(L, 2L)$, drifts with system size according to
\begin{equation}
    T^{\ast}(L, 2L) = T_c + \frac{A}{L^{2/3}} + \cdots,
    \label{eq:Tstar_scaling}
\end{equation}
where $A$ is a non-universal amplitude and the omitted terms decay faster with increasing $L$. This relation is used to extract $T_c$ for the models studied in this work.
\begin{figure}[]
    \centering
    \includegraphics[width=0.48\textwidth]{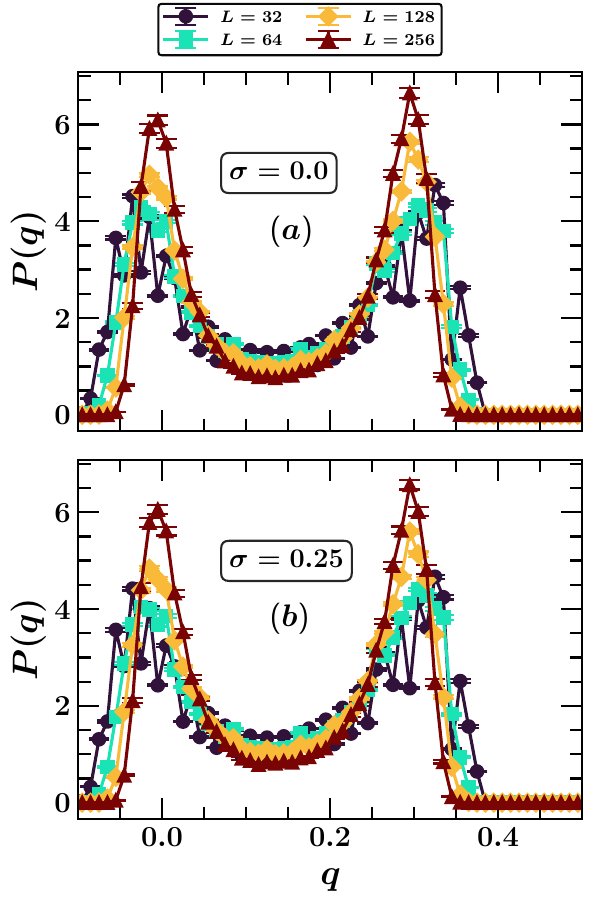}
\caption{
Spin overlap distribution $P(q)$ at the fixed temperature $T = 0.2$
for the fully connected model:
(a) $\sigma = 0$ and (b) $\sigma = 0.25$.
Each panel shows the system-size dependence of the overlap
distribution at the lowest simulated temperature.
}
    \label{fig:spin_overlap_q15_fully_connected}
\end{figure}
\begin{figure}[]
    \centering
    \includegraphics[width=\columnwidth]{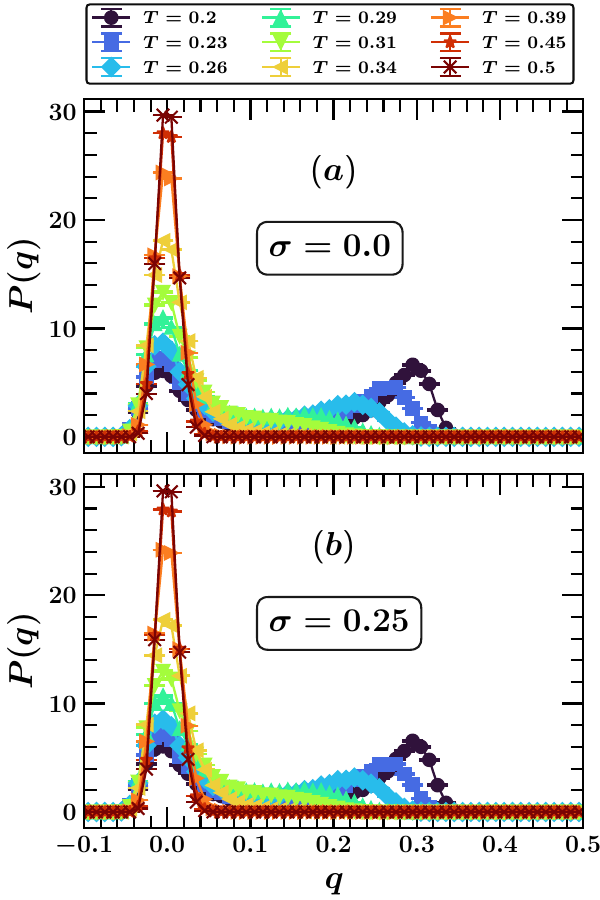}
\caption{
Spin overlap distribution $P(q)$ for the fully connected model
at fixed system size $L = 256$, shown at multiple temperatures
used in the simlulation.
Panels (a) and (b) correspond to $\sigma = 0$ and $\sigma = 0.25$,
respectively.
}
    \label{fig:spin_overlap_256_fully_connected}
\end{figure}

\begin{figure*}
    \centering
    \includegraphics[width=\textwidth]{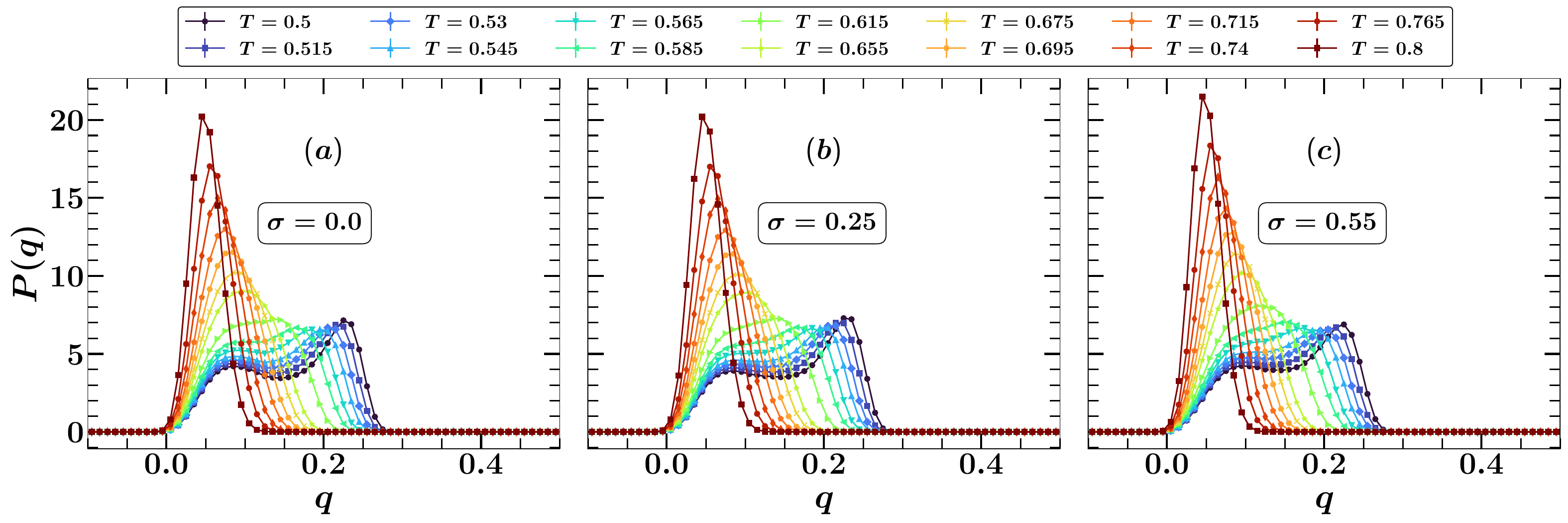}
    \caption{
    Spin overlap distribution $P(q)$ for the power law diluted model at system size
    $L = 1024$.
    From left to right, panels correspond to interaction range parameter
    $\sigma = 0$, $0.25$, and $0.55$.
    Each panel includes data for all simulated temperatures, illustrating the evolution
    of the overlap distribution across different interaction ranges.
    }
    \label{fig:spin_overlap_1024_all_sigma}
\end{figure*}

\begin{figure*}
    \centering
    \includegraphics[width=\textwidth]{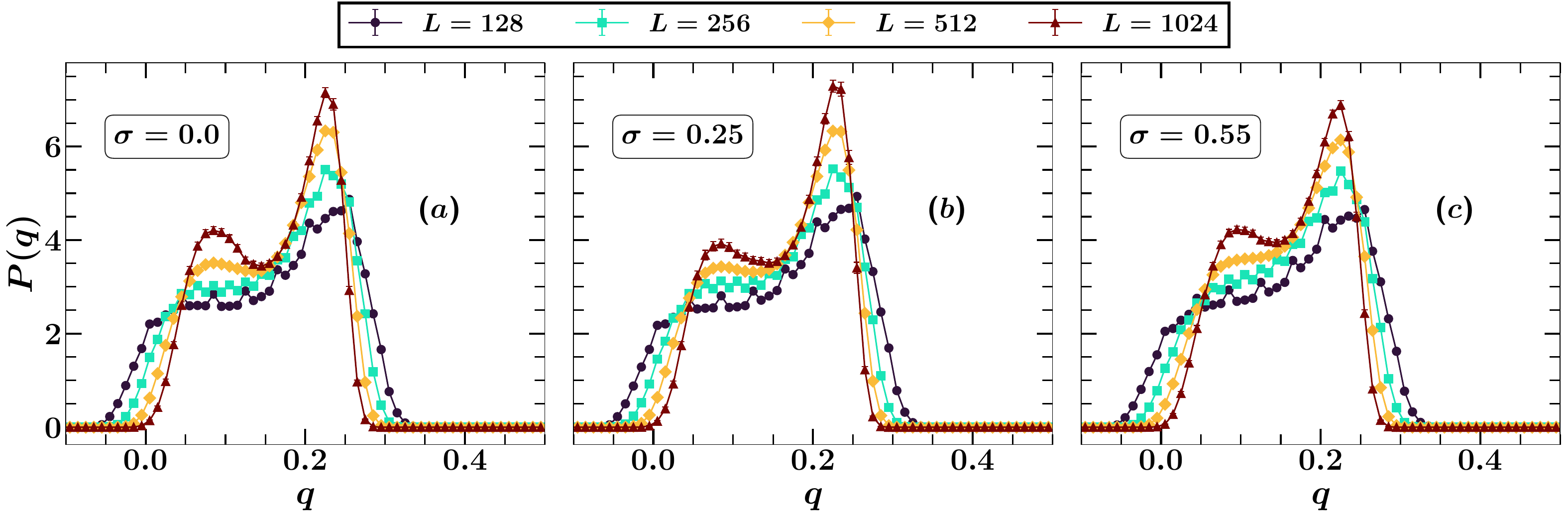}
    \caption{
    Spin overlap distribution $P(q)$ at fixed temperature $T = 0.5$ for the power law
    diluted model.
    From left to right, panels correspond to interaction range exponents
    $\sigma = 0$, $0.25$, and $0.55$.
    Each panel shows the system size dependence of the overlap distribution at a
    particular temperature.
    }
    \label{fig:spin_overlap_q05_all_sigma}
\end{figure*}

\begin{figure*}
    \centering
    \includegraphics[width=\textwidth]{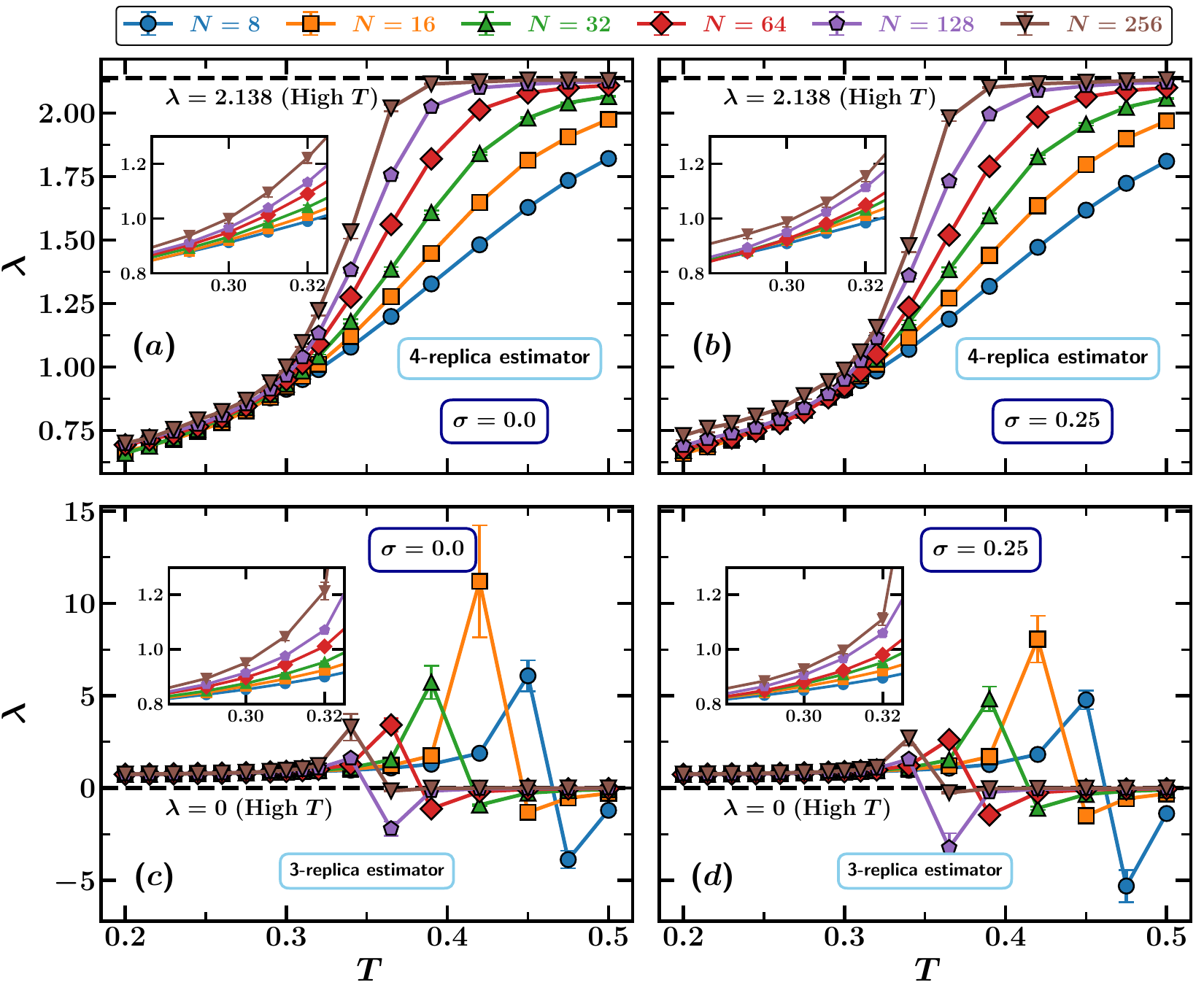}
    \caption{
        Cubic cumulant parameter $\lambda$ as a function of temperature
        for the fully connected model.
        From left to right, panels correspond to $\sigma = 0$ and $0.25$.
        The upper panels show the four replica estimator of $\lambda$,
        while the lower panels display the corresponding three replica
        estimator.
        Insets highlight the behavior of $\lambda$ near the critical
        temperature.
 The black dashed lines denote the high-temperature
limits of $\lambda$. The high-temperature limit is identical for
both $\sigma=0$ and $\sigma=0.25$ within each replica estimator
and agrees excellently with Eq.~(\ref{eq:lambda_highT}).}
    \label{fig:lambda_fully_connected_2sigma}
\end{figure*}
\begin{figure*}
    \centering
    \includegraphics[width=\textwidth]{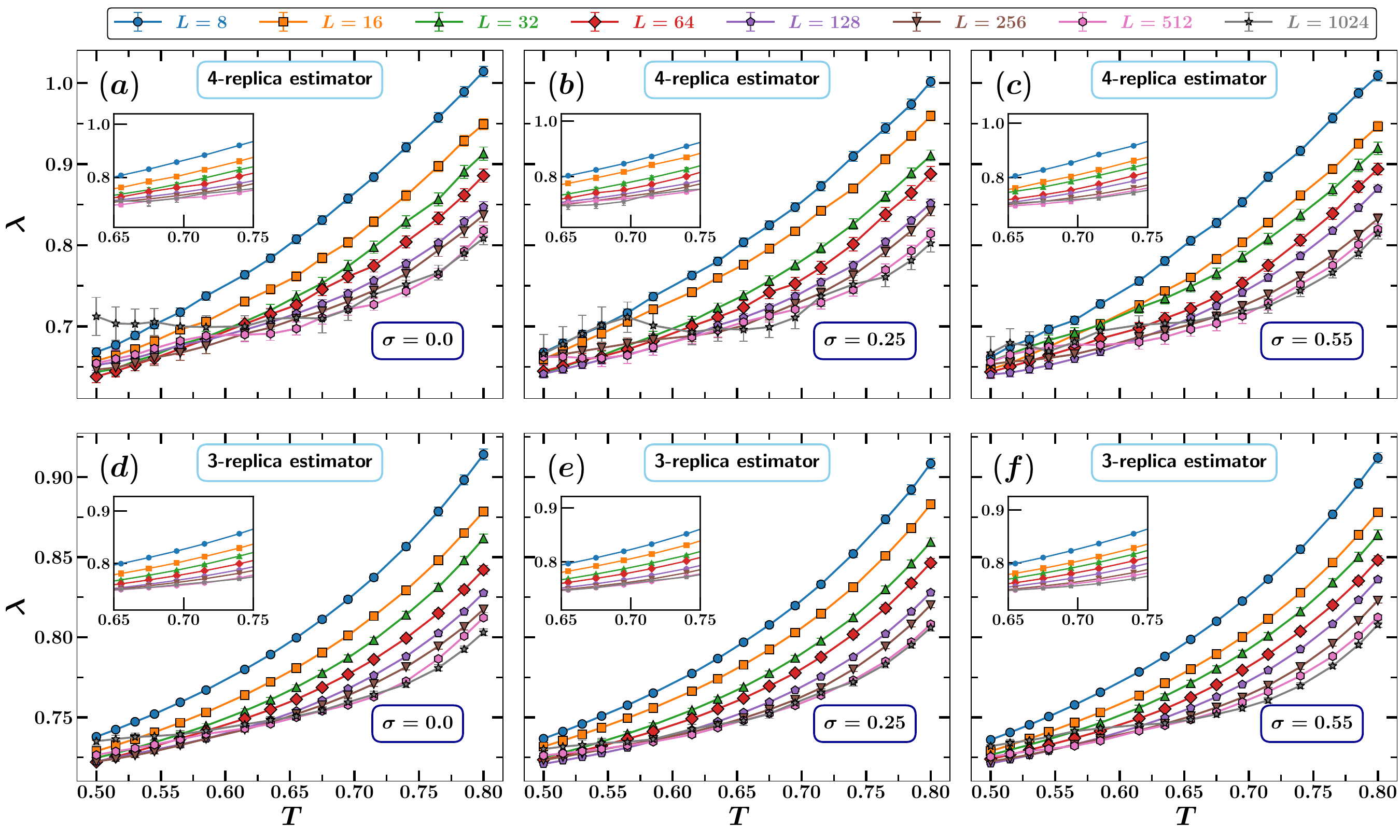}
    \caption{
        Cubic cumulant parameter $\lambda$ as a function of temperature
        for the power law diluted model.
        From left to right, panels correspond to interaction range
        exponents $\sigma = 0$, $0.25$, and $0.55$.
        The upper panels show the four replica estimator of $\lambda$,
        while the lower panels display the corresponding three replica
        estimator.
        Insets highlight the behavior of $\lambda$ in the vicinity of the
        critical temperature.
    }
    \label{fig:lambda_power_law_3sigma}
\end{figure*}
\begin{figure}
    \centering
    \includegraphics[width=\columnwidth]{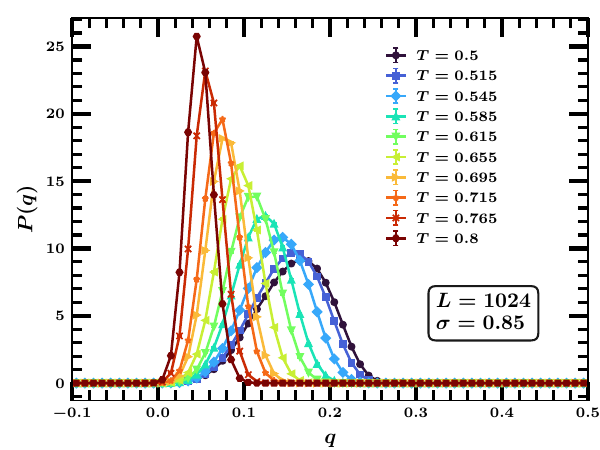}
    \caption{
    Spin overlap distribution $P(q)$ for the diluted case at system size
    $L = 1024$ and interaction range parameter $\sigma = 0.85$.
    }
    \label{fig:spin_overlap_sigma_085_L512}
\end{figure}

\begin{figure}
    \centering
    \includegraphics[width=0.9\columnwidth]{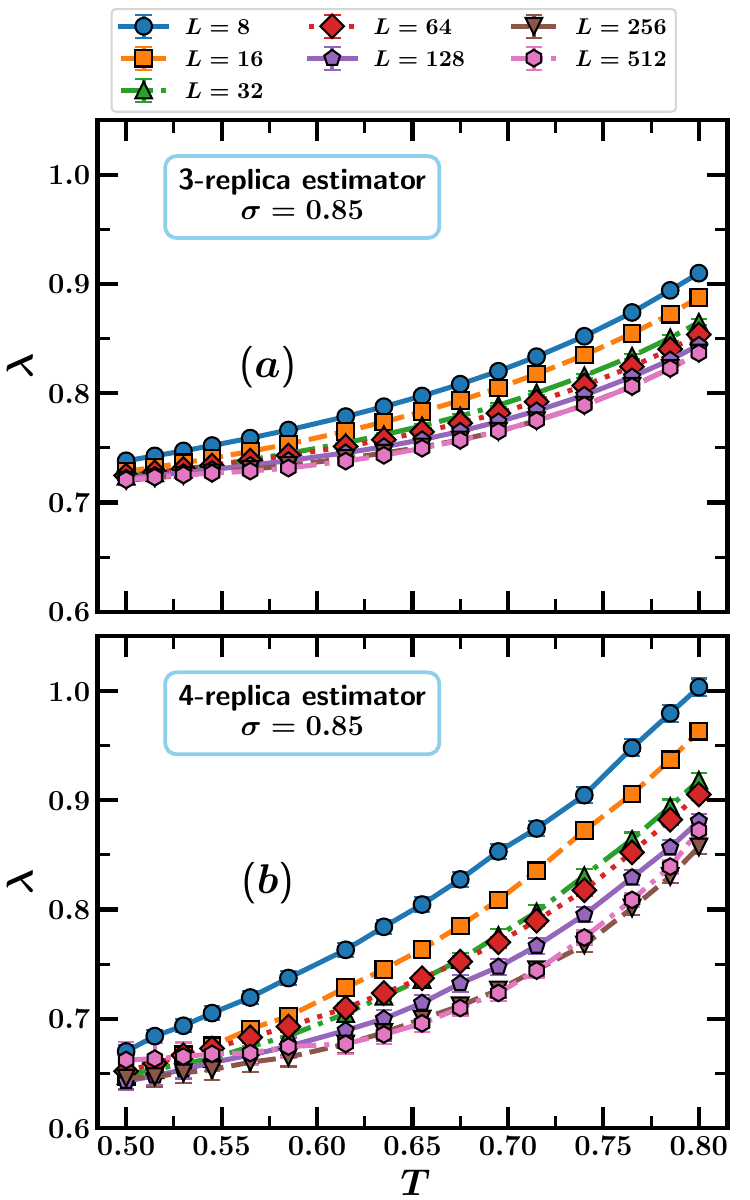}
    \caption{
Temperature variation of the parameter $\lambda$ for the diluted model at $\sigma = 0.85$. Panel (a) corresponds to the three replica estimator, whereas panel (b) shows the corresponding four replica estimator. Data for different system sizes are shown across the entire simulated temperature range.
}
    \label{fig:lambda_sigma_085}
\end{figure}

\begin{figure}
    \centering
    \includegraphics[width=0.48\textwidth]{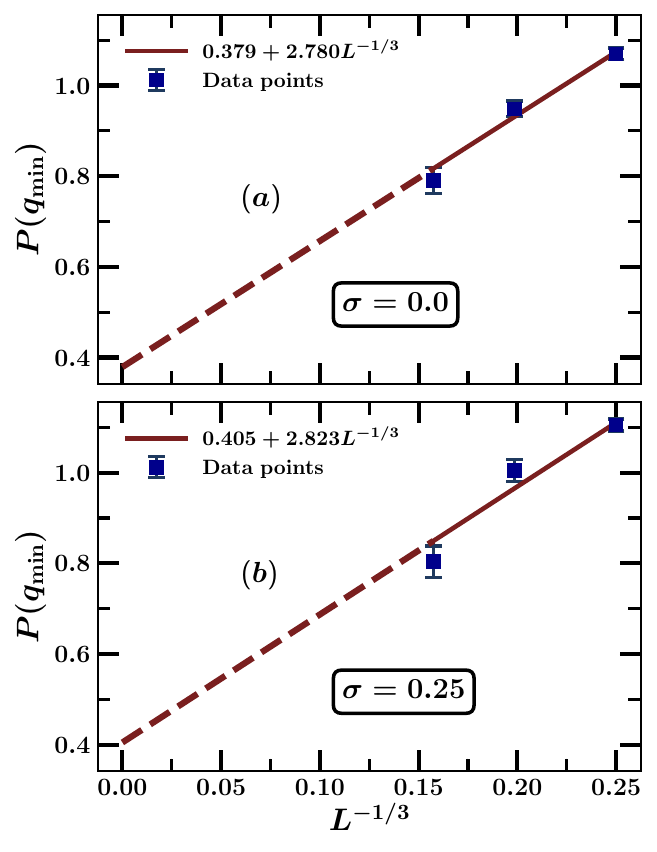}
    \caption{
        Collective finite size scaling analysis of the minimum of the spin overlap
        distribution for the fully connected model at $T = 0.2$.
        The figure shows the behavior of $P(q_{\min})$ as a function of
        $L^{-1/3}$ ($L$ = $64$, $128$ and $256$) for $\sigma = 0$ (top) and $0.25$ (bottom).
        The data are linearly extrapolated in $L^{-1/3}$ to estimate the
        thermodynamic limit value of $P(q_{\min})$.
        The straight lines represent the corresponding fits used to obtain
        the infinite size extrapolation.
    }
    \label{fig:collective_pqmin_fully_connected}
\end{figure}

\subsection{Spin-overlap distribution function}
To characterize the spin glass transition, we introduce an order
parameter defined in terms of the overlap between two independent
replicas of the system evolving under the same realization of the
quenched couplings. 
In the simulations we generate four real replicas and define the replica-overlap matrix \(Q_{ab}\) as~\cite{jyeo,caltagirone}
\begin{equation}
    q \equiv Q_{ab} = \frac{1}{L} \sum_{i=1}^{L} q_{ab}(i),
    \label{eq:Q_ab_rewrite}
\end{equation}
where \(q_{ab}(i)\) is the local overlap at site \(i\).
At each rung there are six distinct pairwise spin products. We
introduce the corresponding composite variables
\(u_{r,i}^{\alpha}\) (\(r=1,\dots,6\)) for replica
\(\alpha\in\{a,b,c,d\}\):
\begin{equation}
\begin{aligned}
    u_{1,i}^{\alpha} &= S_i^{\alpha} T_i^{\alpha}, &
    u_{2,i}^{\alpha} &= S_i^{\alpha} U_i^{\alpha}, &
    u_{3,i}^{\alpha} &= S_i^{\alpha} V_i^{\alpha}, \\[6pt]
    u_{4,i}^{\alpha} &= T_i^{\alpha} U_i^{\alpha}, &
    u_{5,i}^{\alpha} &= T_i^{\alpha} V_i^{\alpha}, &
    u_{6,i}^{\alpha} &= U_i^{\alpha} V_i^{\alpha}.
\end{aligned}
\label{eq:u_definitions_rewrite}
\end{equation}

In terms of these composite variables, the local overlap is written
in a compact and symmetric form as
\begin{equation}
    q_{ab}(i)
    = \frac{1}{M^{2}}
      \sum_{r=1}^{6}
      u_{r,i}^{a}\,u_{r,i}^{b},
    \qquad (M=4),
    \label{eq:q_ab_rewrite}
\end{equation}
so that \(Q_{ab}\) corresponds to the spatial average of the local overlap.
This definition reflects the structure of the Hamiltonian, where the interactions are written in terms of pairwise spin products.
A central observable in the spin-glass phase is the distribution of the order parameter $P(q)$.
The distribution \(P(q)\) is obtained by sampling equilibrium configurations at temperature \(T\) for a fixed disorder realization and subsequently averaging over disorder samples.
Its shape provides direct information about the organization of states in the spin-glass phase.
In a replica-symmetric phase, $P(q)$ becomes sharply
peaked around a single value (typically near zero in the
paramagnetic phase). In contrast, in a phase with replica symmetry breaking (RSB), the overlap distribution develops a nontrivial structure. For a one-step RSB (1RSB) scenario, two
well-separated peaks emerge below the transition temperature
\(T_c\), reflecting the coexistence of distinct pure states. This bimodal structure is reminiscent of the order-parameter distribution at a first-order transition and signals a discontinuous reorganization of phase space.
\subsection{Study of $\lambda$-parameter}
The $\lambda$-parameter plays a central role in characterizing the nature of the glassy transition within the replicated Gibbs free energy framework. It is defined as the ratio of the two independent cubic coefficients, $\omega_2$ and $\omega_1$, appearing in the expansion of the replicated free energy~\cite{Parisi_ratio_2013,PhysRevLett.108.085702}:
\begin{equation}
\label{eq:lam_ratio_def}
\lambda = \frac{\omega_2}{\omega_1}.
\end{equation}

The coefficients $\omega_1$ and $\omega_2$ can be expressed as linear combinations of eight independent cubic cumulants $W_i$~\cite{Parisi_ratio_2013, martin_mayor,PhysRevE.105.054106}. Explicitly,
\begin{equation}
\label{eq:omega_def}
\begin{aligned}
\omega_1 &= W_1 - 3W_5 + 3W_7 - W_8, \\[4pt]
\omega_2 &= \tfrac{1}{2}W_2 - 3W_3 + \tfrac{3}{2}W_4
          + 3W_5 + 2W_6 - 6W_7 + 2W_8 .
\end{aligned}
\end{equation}

These cumulants $W_i$ quantify different types of three-point correlations of the overlap fluctuations. They are defined in terms of the fluctuations of the overlap matrix $Q_{ab}$ as
\begin{equation}
\begin{aligned}
W_1 &\equiv L^2 \, \overline{\big\langle \delta \tilde{Q}_{ab}\,
\delta \tilde{Q}_{bc}\,\delta \tilde{Q}_{ca} \big\rangle}, \\
W_2 &\equiv L^2 \, \overline{\big\langle (\delta \tilde{Q}_{ab})^3 \big\rangle}, \\
W_3 &\equiv L^2 \, \overline{\big\langle (\delta \tilde{Q}_{ab})^2\,
\delta \tilde{Q}_{ac} \big\rangle}, \\
W_4 &\equiv L^2 \, \overline{\big\langle (\delta \tilde{Q}_{ab})^2\,
\delta \tilde{Q}_{cd} \big\rangle}, \\
W_5 &\equiv L^2 \, \overline{\big\langle \delta \tilde{Q}_{ab}\,
\delta \tilde{Q}_{ac}\,\delta \tilde{Q}_{bd} \big\rangle}, \\
W_6 &\equiv L^2 \, \overline{\big\langle \delta \tilde{Q}_{ab}\,
\delta \tilde{Q}_{ac}\,\delta \tilde{Q}_{ad} \big\rangle}, \\
W_7 &\equiv L^2 \, \overline{\big\langle \delta \tilde{Q}_{ab}\,
\delta \tilde{Q}_{ac}\,\delta \tilde{Q}_{de} \big\rangle}, \\
W_8 &\equiv L^2 \, \overline{\big\langle \delta \tilde{Q}_{ab}\,
\delta \tilde{Q}_{cd}\,\delta \tilde{Q}_{ef} \big\rangle}.
\end{aligned}
\end{equation}
Here $\langle \cdots \rangle$ denotes the thermal average, while the overline represents the disorder average.
The overlap fluctuations are defined with respect to the mean overlap as
\begin{equation}
    \delta \tilde{Q}_{ab}
    = \frac{1}{L M^{2}} 
    \sum_{i=1}^{L}
    \left(
        \sum_{r=1}^{6}
        u_{r,i}^{a} \, u_{r,i}^{b}
    \right) - \langle Q_{ab} \rangle,
    \label{eq:deltaQ}
\end{equation}
which captures both thermal and sample-to-sample fluctuations of the replicated overlap field.
To compute the cubic couplings $\omega_{1}$ and $\omega_{2}$, one would ideally simulate six real replicas of the system, meaning six statistically independent configurations evolving under the same disorder realization. In practice, these coefficients can be rewritten in terms of alternative estimators that require only three or four real replicas. Within the replica-symmetric (RS) description~\cite{Parisi_ratio_2013}, the three and four replica estimators do not generally reproduce the exact definitions of $\omega_{1}$ and $\omega_{2}$; nevertheless, the differences vanish as the critical point is approached,
\[
\omega_i-\omega_i^{(3)}=O(|T-T_c|), \qquad
\omega_i-\omega_i^{(4)}=O(|T-T_c|^3),
\]
so that all estimators converge to the same critical value at $T_c$~\citep{PhysRevE.105.054106}. Since our primary interest is the critical value of $\lambda$, and in particular whether $\lambda$ is greater or smaller than unity, the three- and four-replica estimators are sufficient for the purposes of this work. This property also provides a useful consistency check of the RS scenario by directly comparing the results obtained from the three- and four-replica constructions \cite{Parisi_ratio_2013,PhysRevE.105.054106,martin_mayor}.
  Additional insight is provided by the behaviour of the parameter $\lambda$, which is defined as the ratio of $\omega_{1}$ and $\omega_{2}$.  Within RS field theory, a continuous transition requires $0 \leq \lambda \leq 1$, while values $\lambda > 1$ are indicative of a weakly first-order transition mechanism~\cite{PhysRevE.101.042114}. Moreover, the same parameter $\lambda$ governs the mean-field expectations for several equilibrium and non-equilibrium dynamical critical exponents, making it a particularly interesting quantity. 
We emphasize that the criterion $\lambda<1$ is assessed at the transition temperature $T_c$
(see Fig.~\ref{fig:lambda_fully_connected_2sigma}). In the fully
connected mean-field model, however, a 1RSB transition,
if present, is expected to occur at a slightly higher temperature
$T_{1\mathrm{RSB}}$. Although $\lambda$ increases with temperature,
our numerical estimates indicate $\lambda<1$ in the vicinity of
$T_c$. Since the temperature difference
$T_{1\mathrm{RSB}}-T_c$ is expected to be small, the present data
do not allow us to determine with certainty whether $\lambda$
remains below unity at $T_{1\mathrm{RSB}}$. So, together
with the absence of numerical signatures of a thermodynamic 1RSB
transition, the observed behavior of $\lambda$ provides supporting
evidence against a stable 1RSB phase in this model.

In Appendix~\ref{appendix:B}, we have shown the high temperature computation of $\lambda$-parameter using three and four replica estimators.
\section{Results}
\label{sec:Results}

In this section, we present the numerical results for both the power-law fully connected
and the power-law diluted models. Our analysis focuses on the behavior
of the spin glass susceptibility and its finite-size scaling properties, which
are common to both models. In addition to this, we further examine the spin overlap distribution and cubic cumulants in order to characterize the
nature of the spin glass phase.

We begin with the fully connected model. Fig.~\ref{fig:fc_collective_susceptibility}
shows the rescaled wave-vector dependent spin glass susceptibility for $\sigma = 0$ and $\sigma=0.25$ as a function of temperature across different system sizes. 
A clear crossing behavior is observed, providing strong evidence for a finite-temperature spin glass transition.
To estimate the critical temperature in the thermodynamic limit, we perform a finite-size scaling analysis based on the intersections of the rescaled susceptibility curves. Using the scaling form given in Eq.~\eqref{eq:Tstar_scaling}, the intersection temperatures $T^{*}(L,2L)$ are extracted and plotted as a function of $1/L^{0.67}$, as shown in Fig.~\ref{fig:collective_intersection_fully_connected}.
The transition temperature is obtained from the crossing between the linear extrapolation of the intersection points. This analysis provides a well-defined estimate of the critical temperature $T_c = 0.320(7)$ for $\sigma=0$ and $T_c = 0.322(6)$ for $\sigma=0.25$ for the fully connected model.
Fig.~\ref{fig:diluted_sus_and_collapse_collective}
summarizes the behavior of the rescaled spin glass susceptibility for three values of $\sigma$, which are  $0$, $0.25$,
and $0.55$ for the power law diluted model. The cases $\sigma = 0$ and $\sigma=0.25$ lie in the non-extensive regime, whereas $\sigma = 0.55$ belongs to the mean-field regime. This choice allows us to systematically investigate the evolution of the critical
behavior as the interaction range is varied. The top row of Fig.~\ref{fig:diluted_sus_and_collapse_collective} shows the
temperature dependence of the rescaled susceptibility for system sizes ranging from $L=8$ up to $L=1024$.

For all three values of $\sigma$, the susceptibility curves exhibit approximate
crossings, indicating the presence of a finite-temperature spin glass transition. The crossing points display a systematic drift with increasing system size, reflecting stronger finite-size effects in the diluted model.
To account for this behavior, we perform a finite-size scaling analysis based on the intersections of the susceptibility curves, as shown in Fig.~\ref{fig:collective_intersection_power_law}.
The intersection temperatures are analyzed using a bootstrap resampling method, allowing for a reliable estimation of both the critical temperature
and its statistical uncertainty.
The resulting extrapolated critical temperatures are then used to rescale the
susceptibility data according to the finite-size scaling form given in
Eq.~\eqref{eq:chi_scaling_meanfield}.
The bottom row of Fig.~\ref{fig:diluted_sus_and_collapse_collective} shows the
corresponding data collapse for each value of $\sigma$.
In all cases, we observe a satisfactory collapse across the full range of system
sizes, providing strong support for the validity of the scaling relation.
From this analysis, we obtain the following estimates for the critical
temperature:
$T_c = 0.701(4)$ for $\sigma = 0$,
$T_c = 0.695(4)$ for $\sigma = 0.25$, and
$T_c = 0.673(4)$ for $\sigma = 0.55$.
Fig.~\ref{fig:combined_sigma_085} presents the finite-size behavior of the diluted model in the non-mean field regime at $\sigma = 0.85$, which shows the rescaled spin-glass susceptibility. It is evident that the curves for different system sizes do not exhibit any clear crossing within the simulated temperature range. This absence of a crossing supports that no thermodynamic phase transition occurs in this regime. Instead, the results indicate a progressive growth of characteristic length scales as the temperature is lowered.

Having established the location of the transition, we next examine the
nature of the low-temperature phase in both the
models. Figures~\ref{fig:spin_overlap_q15_fully_connected} and
\ref{fig:spin_overlap_256_fully_connected} show the behavior of the
spin overlap distribution $P(q)$ for the fully connected model.  In
Fig.~\ref{fig:spin_overlap_q15_fully_connected}, we present $P(q)$ at
the fixed temperature $T=0.2$ for several system sizes, suggesting the
system size dependence of the overlap distribution for $\sigma=0$ and
$0.25$.  To further examine the temperature dependence,
Fig.~\ref{fig:spin_overlap_256_fully_connected} shows $P(q)$ for the
largest system size $L=256$ at several temperatures used in the
simulations. These results allow us to examine how the overlap
distribution evolves with both system size and temperature, providing
understanding of the nature of the low-temperature phase of the fully
connected model. Fig.~\ref{fig:spin_overlap_1024_all_sigma} shows the
spin overlap distribution function $P(q)$ for the diluted model for
$L=1024$ at all simulated temperatures.  At temperatures well below
the critical temperature, the overlap distribution exhibits broad
support and nontrivial structure, characteristic of a full replica
symmetry breaking (FRSB) phase.  In contrast, at high temperatures ($T
> T_c$), the distribution becomes sharply peaked around $q=0$,
consistent with paramagnetic behavior.  Further insight is obtained in
Fig.~\ref{fig:spin_overlap_q05_all_sigma}, where $P(q)$ is shown at
the lowest temperature $T=0.5$ for the four largest system sizes and
for all three values of $\sigma$ for the diluted case.  The systematic
evolution of the overlap distribution with increasing system size
provides additional evidence for the persistence of an FRSB phase in
the thermodynamic limit.

In Fig.~\ref{fig:lambda_fully_connected_2sigma} and
\ref{fig:lambda_power_law_3sigma} respectively, the behavior of the
$\lambda$ parameter for the fully connected and diluted model is
shown. $\lambda$-parameter is computed using both three (See
Eq.~\eqref{three_rep_estimator}) and four (See
Eq.~\eqref{four_rep_estimator}) replica estimators.  For all values of
$\sigma$, $\lambda$ remains nearly less than unity in the vicinity of
the critical temperature except for large system sizes for the fully
connected case as shown in insets of
Fig.~\ref{fig:lambda_fully_connected_2sigma}. 

For the diluted case in
Fig.~\ref{fig:lambda_power_law_3sigma}, $\lambda$ remains strictly
less than unity for both three and four replica estimators in the
vicinity of criticality, as highlighted in the inset panels.  This
behavior is consistent with a transition from a replica symmetric
phase at high temperatures to a full replica symmetry breaking phase
below $T_c$, and in close agreement with the conclusions drawn from
the overlap distribution analysis.  
In Fig.~\ref{fig:spin_overlap_sigma_085_L512}, we present the behavior of the spin overlap distribution $P(q)$ for the diluted model at $\sigma = 0.85$ and system size $L = 1024$ across a range of temperatures. The results show no indication of the characteristic two-peak structure associated with a one-step replica symmetry breaking (1RSB) transition for a physical  dimension like $d = 3$.
To further support this observation, we compute the $\lambda$-parameter for $\sigma = 0.85$, as shown in Fig.~\ref{fig:lambda_sigma_085}. Panel (a) displays the three-replica estimator, while panel (b) presents the corresponding four-replica estimator. In both cases, $\lambda$ remains consistently below unity over the entire temperature range studied.
These findings provide strong evidence against the presence of a 1RSB-like or thermodynamic phase transition in this regime. Instead, the results support a scenario characterized by a continuous growth of correlation length scales as the temperature is lowered, without the emergence of a distinct phase transition.
In Fig.~\ref{fig:collective_pqmin_fully_connected}, we examine the
behavior of the minimum of the spin-overlap distribution,
$P(q_{\min})$, for the fully connected model for $\sigma=0$ and $0.25$ in panel $(a)$ and $(b)$ respectively using the three largest
system sizes accessible in our simulations. This analysis is aimed at
testing whether $P(q_{\min})$ vanishes in the thermodynamic limit, as
expected for a one-step replica symmetry breaking (1RSB) transition.

Our finite-size scaling analysis indicates that $P(q_{\min})$ remains
finite upon extrapolation to the thermodynamic limit. This behavior
provides no evidence for a 1RSB transition and, consequently, does not
support the presence of a Gardner transition within the range of
system sizes and parameters explored in this study.

However, we emphasize that the observed behavior at least for the fully connected case  at $\sigma = 0$  is dominated 
by finite-size effects. This interpretation is consistent with our
analysis of the $\lambda$-parameter, which indicates that much larger
system sizes are likely required to resolve its behavior for the thermodynamic limit. In particular, a crossing to $\lambda >
1$ which we would  expect at larger sizes  for the fully connected model at $\sigma= 0$ would signal the onset of a discontinuous (1RSB)
transition followed by a transition to a full replica symmetry
breaking (FRSB) Gardner phase at lower temperatures.

\section{Summary and Conclusion}
\label{sec:summary and conclusion}

In this work, we have investigated both the fully connected model in
the non-extensive regime and the power-law diluted model across all
three relevant regimes: non-extensive ($\sigma = 0$ and $0.25$),
mean-field ($\sigma = 0.55$), and non-mean-field ($\sigma = 0.85$).
To characterize the phase transition and determine the critical
temperature, we analyzed the spin-glass susceptibility together with
the spin-overlap distribution function.

For the fully connected model, the finite-size scaling analysis shown
in Fig.~\ref{fig:collective_intersection_fully_connected} yields a
transition temperature $T_c \approx 0.320(7)$, in good agreement with
the theoretical predictions of Yeo and Moore~\cite{jyeo}.
Furthermore, our results support the expectation that different values
of $\sigma$ in the non-extensive regime yield the same transition
temperature, consistent with the arguments of
Mori~\cite{Mori_2011}.

The statistical uncertainties were estimated using bootstrap
resampling, while the fitting procedures were carried out using the
\texttt{scipy.curve\_fit} routine. The largest system sizes accessible
in our simulations do not exhibit clear signatures of a first-order
transition; both the behavior of the $\lambda$-ratio, which remains
below unity, and the form of the overlap distribution function
consistently support this observation. In particular, the $\lambda <
1$ behavior (near criticality) points toward a continuous
transition~\cite{PhysRevE.101.042114}, while the overlap distribution
function lacks a well-resolved two-peak structure and instead appears
as a broadened, nearly continuous distribution.

A similar behavior is observed in the diluted model for $\sigma = 0$,
$0.25$, and $0.55$, for which we estimate the critical temperature to
be $T_c \approx 0.72$, consistent with the independent calculation
presented in Appendix~\ref{appendix:C}. For the non-mean-field case
$\sigma = 0.85$, however, neither the rescaled susceptibility nor the
correlation length exhibits any clear crossing within the simulated
temperature range. Consistently, the $\lambda$-parameter remains below
unity for both the three-replica and four-replica estimators, and the
spin-overlap distribution does not show any evidence of the
characteristic two-peak structure associated with a 1RSB transition.
Instead, the results indicate only a progressive growth of length
scales as the temperature is reduced.
Our results highlight the important role of finite-size effects in
shaping the observed behavior. In the fully connected model, finite
system sizes effectively renormalize the coefficients $\omega_1$ and
$\omega_2$ in the replicated Gibbs free energy expansion, leading to
an effective ratio $\lambda < 1$. This may suppress the appearance of a
one-step replica symmetry breaking (1RSB) transition in the accessible
numerical regime. Although higher-order contributions are naturally
included in the simulations, one possible effect  of finite size is to
shift these coefficients away from their asymptotic values obtained in
the steepest descent approximation~\cite{jyeo}. We note that the same finite-size effects could, in principle, also influence the value of $\lambda$ and the shape of the overlap distribution; nevertheless, neither observable shows any systematic trend toward the expected 1RSB signatures within the accessible system sizes.  Consequently, we cannot exclude the possibility that simulations at substantially larger system sizes might eventually reveal the expected 1RSB transition. Beyond finite-size effects, however, it is well established that the dynamical transition is wiped out in finite spatial dimensions, and a similar fate may hold for the static 1RSB transition itself. Indeed, the features of the ordered phase below $T_c$ do not exhibit the character predicted by Rizzo's replica field theory~\cite{PhysRevE.88.032135}, which would be expected if a 1RSB transition actually occurred. We thus tentatively suggest that the thermodynamic 1RSB transition does not exist in finite dimensions.

For the diluted model, we likewise do not observe clear evidence of a
1RSB transition within the range of system sizes studied. In
particular, for $\sigma = 0.85$, which approximately corresponds to
the three-dimensional regime, our numerical results are consistent with
the absence of both 1RSB and FRSB transitions over the temperature range
investigated. While we cannot completely rule out alternative crossover
scenarios on the basis of the presently accessible system sizes, the
available numerical evidence favors this scenario. Consequently,
our results suggest that the Kauzmann temperature may be zero in the
corresponding three-dimensional system.

\section*{Acknowledgments}
We gratefully acknowledge the High Performance Computing (HPC) facility at IISER Bhopal and the PARAM SHIVAY supercomputing facility at IIT (BHU), Varanasi, for providing the computational resources used to perform the large-scale numerical simulations reported in this work. P.~G. acknowledges IISER Bhopal, India, for financial support through a PhD fellowship. We would like to thank Tommaso Rizzo for helpful discussions and useful exchanges.
\appendix
\section{Hamiltonian in expanded form}

In this appendix, we present the explicit expanded form of the Hamiltonian introduced in Eq.~\eqref{Hamil_compact}. For the $M=4$, $p=4$ model, the interaction between two neighboring sites (rungs) $i$ and $j$ involves pairwise combinations of spins within each rung, resulting in a total of 36 independent four-spin interaction terms.

\begin{widetext}
\begin{equation}
\label{eq:Hamiltonian_M4p4_pair}
\begin{aligned}
\mathcal{H} = -\sum_{\langle ij \rangle} \Bigg[
& S_{i} T_{i}
\Big(
J_{ij}^{(1)} S_{j} T_{j}
+ J_{ij}^{(2)} S_{j} U_{j}
+ J_{ij}^{(3)} S_{j} V_{j}
+ J_{ij}^{(4)} T_{j} U_{j}
+ J_{ij}^{(5)} T_{j} V_{j}
+ J_{ij}^{(6)} U_{j} V_{j}
\Big) \\
& + S_{i} U_{i}
\Big(
J_{ij}^{(7)} S_{j} T_{j}
+ J_{ij}^{(8)} S_{j} U_{j}
+ J_{ij}^{(9)} S_{j} V_{j}
+ J_{ij}^{(10)} T_{j} U_{j}
+ J_{ij}^{(11)} T_{j} V_{j}
+ J_{ij}^{(12)} U_{j} V_{j}
\Big) \\
& + S_{i} V_{i}
\Big(
J_{ij}^{(13)} S_{j} T_{j}
+ J_{ij}^{(14)} S_{j} U_{j}
+ J_{ij}^{(15)} S_{j} V_{j}
+ J_{ij}^{(16)} T_{j} U_{j}
+ J_{ij}^{(17)} T_{j} V_{j}
+ J_{ij}^{(18)} U_{j} V_{j}
\Big) \\
& + T_{i} U_{i}
\Big(
J_{ij}^{(19)} S_{j} T_{j}
+ J_{ij}^{(20)} S_{j} U_{j}
+ J_{ij}^{(21)} S_{j} V_{j}
+ J_{ij}^{(22)} T_{j} U_{j}
+ J_{ij}^{(23)} T_{j} V_{j}
+ J_{ij}^{(24)} U_{j} V_{j}
\Big) \\
& + T_{i} V_{i}
\Big(
J_{ij}^{(25)} S_{j} T_{j}
+ J_{ij}^{(26)} S_{j} U_{j}
+ J_{ij}^{(27)} S_{j} V_{j}
+ J_{ij}^{(28)} T_{j} U_{j}
+ J_{ij}^{(29)} T_{j} V_{j}
+ J_{ij}^{(30)} U_{j} V_{j}
\Big) \\
& + U_{i} V_{i}
\Big(
J_{ij}^{(31)} S_{j} T_{j}
+ J_{ij}^{(32)} S_{j} U_{j}
+ J_{ij}^{(33)} S_{j} V_{j}
+ J_{ij}^{(34)} T_{j} U_{j}
+ J_{ij}^{(35)} T_{j} V_{j}
+ J_{ij}^{(36)} U_{j} V_{j}
\Big)
\Bigg].
\end{aligned}
\end{equation}
\end{widetext}

The structure of Eq.~\eqref{eq:Hamiltonian_M4p4_pair} reflects the fact that each rung contains $M=4$ spins, from which $\binom{4}{2} = 6$ distinct spin pairs can be formed. The interaction between two neighboring rungs $i$ and $j$ therefore involves all possible pairwise combinations of these spin pairs, resulting in a total of $(\binom{4}{2})^2 = 36$ four-spin interaction terms.
Each term corresponds to the interaction between a pair of spins on rung $i$ and a pair of spins on rung $j$, as illustrated schematically in Fig.~\ref{fig:schematic}. The corresponding coupling constants $J_{ij}^{(k)}$ are independent quenched random variables drawn from a Gaussian distribution, ensuring statistical independence among the different interaction channels.
\section{Cubic cumulants in the high temperature limit}\label{appendix:B}
In this appendix, we analyze the behavior of the cubic cumulants in the high temperature phase and evaluate the corresponding estimates of the parameter
\begin{equation}
\lambda = \frac{\omega_2}{\omega_1}.
\end{equation}
We consider both the three replica and four replica estimators of the cubic cumulants, denoted collectively by \(\mathcal{W}\)'s.
%
%
%

The three replica estimator is given by~\cite{Parisi_ratio_2013}
\begin{equation}
\label{three_rep_estimator}
\begin{aligned}
\omega_1^{(3)} &= \frac{11}{30} W_1 - \frac{2}{15} W_2, \\
\omega_2^{(3)} &= \frac{4}{15} W_1 - \frac{1}{15} W_2,
\end{aligned}
\end{equation}
while the four replica estimator reads
\begin{equation}
\label{four_rep_estimator}
\begin{aligned}
\omega_1^{(4)} &= \frac{23}{30} W_1 + \frac{1}{20} W_2 - \frac{3}{5} W_3
+ \frac{9}{20} W_4 - \frac{6}{5} W_5 + \frac{1}{2} W_6, \\
\omega_2^{(4)} &= \frac{7}{15} W_1 + \frac{2}{5} W_2 - \frac{9}{5} W_3
+ \frac{3}{5} W_4 - \frac{3}{5} W_5 + W_6 .
\end{aligned}
\end{equation}

The fluctuation of the overlap matrix is defined as
\begin{equation}
\delta \tilde{Q}_{ab} \equiv Q_{ab} - \langle Q_{ab} \rangle,
\end{equation}
with
\begin{equation}
\label{eq:Q_and_qab_reln}
Q_{ab} = \frac{1}{L}\sum_{i=1}^{L} q_{ab}(i).
\end{equation}
The local overlap at site \(i\) between replicas \(a\) and \(b\) is
\begin{equation}
q_{ab}(i) = \frac{1}{M^2}\sum_{r=1}^{6} u_{r,i}^a\,u_{r,i}^b,
\end{equation}
where the six composite variables are
\begin{equation}
\begin{aligned}
u_{1,i}^{\alpha} &= S_i^{\alpha} T_i^{\alpha}, &
u_{2,i}^{\alpha} &= S_i^{\alpha} U_i^{\alpha}, &
u_{3,i}^{\alpha} &= S_i^{\alpha} V_i^{\alpha}, \\
u_{4,i}^{\alpha} &= T_i^{\alpha} U_i^{\alpha}, &
u_{5,i}^{\alpha} &= T_i^{\alpha} V_i^{\alpha}, &
u_{6,i}^{\alpha} &= U_i^{\alpha} V_i^{\alpha}.
\end{aligned}
\end{equation}

We now evaluate the leading non-vanishing contribution to
\begin{equation}
W_1 = L^2 \, \overline{\big\langle
\delta \tilde{Q}_{ab}\,\delta \tilde{Q}_{bc}\,\delta \tilde{Q}_{ca}
\big\rangle}.
\end{equation}
In the high temperature limit the average overlap vanishes
\begin{equation}
\langle Q_{ab} \rangle = 0,
\end{equation}
implying \(\langle u_{p,i} \rangle = 0\) for all \(p\). Consequently, \(\delta \tilde{Q}_{ab} = Q_{ab}\). Thus
\begin{equation}
W_1 = L^2 \, \overline{\big\langle Q_{ab} Q_{bc} Q_{ca} \big\rangle}.
\end{equation}

At high temperature only terms with coinciding site $(i=j=k)$ indices contribute. Using the compact representation of the local overlap, one obtains from Eq.~\eqref{eq:Q_and_qab_reln}
\begin{equation}
q_{ab}(i)\,q_{bc}(i)\,q_{ca}(i)
= \frac{1}{M^6}\sum_{p,q,r=1}^{6}
\big(u_{p,i}^a u_{p,i}^b\big)
\big(u_{q,i}^b u_{q,i}^c\big)
\big(u_{r,i}^c u_{r,i}^a\big).
\end{equation}
The triple sum naturally separates into contributions where all indices are equal, two are equal, or all are distinct. A direct evaluation shows that only the fully contracted terms survive in the high temperature limit~\cite{martin_mayor}.

As a result, among all cubic cumulants only \(W_1\) and \(W_2\) are non zero, with
\begin{equation}
\label{eq:W1_W2_at_hight}
\begin{aligned}
{W}_1 = \frac{6}{M^6}, \, \, 
{W}_2 = \frac{24}{M^6}.
\end{aligned}
\end{equation}

Substituting these values into Eqs.~\eqref{three_rep_estimator} and
\eqref{four_rep_estimator}, we obtain for the high temperature limit
\begin{equation}
\begin{aligned}
\lambda^{(3)} &= \frac{\omega_2^{(3)}}{\omega_1^{(3)}} = 0, \\
\lambda^{(4)} &= \frac{\omega_2^{(4)}}{\omega_1^{(4)}} = 2.138 .
\end{aligned}
\label{eq:lambda_highT}
\end{equation}
\onecolumngrid
\section{Analytical study for the diluted case}
\label{appendix:C}
Here we consider dilute $M$-$p$ spin glasses following
the prescription given by Viana and Bray~\cite{Viana_Bray_1985}.
The $M$ spins at a rung $x$ are denoted by $S_i(x)$, $i=1,2,\cdots,M$. The Hamiltonian for the $p=4$ model is 
\begin{align}
H=-\sum_{(x, y)}\left[ \sum_{i_1<i_2}^M\sum_{j_1<j_2}^M J^{(i_1,i_2),(j_1,j_2)}_{x,y} S_{i_1}(x)S_{i_2}(x)
S_{j_1}(y)S_{j_2}(y)
\right] ,
\end{align}
where the sum is over all pairs of rungs $(x,y)$. 
In the Viana-Bray model~\cite{Viana_Bray_1985}, 
only a fraction $z/L$ bonds are
present, and 
the distribution for  each $J^{(i_1,i_2),(j_1,j_2)}_{x,y}$ is given by
\begin{align}
    P( J^{(i_1,i_2),(j_1,j_2)}_{x,y})=\frac z L P_z(J^{(i_1,i_2),(j_1,j_2)}_{x,y})+
    \left(1-\frac z L\right)\delta(J^{(i_1,i_2),(j_1,j_2)}_{x,y}).
\end{align}
where $P_z$ may be any suitable distribution.
We denote the averages over $P$ and $P_z$
by $\langle ~ \rangle_J$ and $\langle~\rangle$, respectively.

The replicated partition function averaged over the disorder is given by
\begin{align}
\langle Z^n\rangle_J=&\mathrm{Tr} \left\langle \exp \Bigg[\beta 
\sum_{(x, y)} \sum_{i_1<i_2}^M\sum_{j_1<j_2}^M J^{(i_1,i_2),(j_1,j_2)}_{x,y} \sum_a S^a_{i_1}(x)S^a_{i_2}(x)
S^a_{j_1}(y)S^a_{j_2}(y)
\Bigg]\right\rangle_J \nonumber \\
=& \mathrm{Tr} \prod_{(x,y)}\prod_{i_1<i_2}\prod_{j_1<j_2}\Bigg[
\left( 1-\frac z L\right)+\frac{z}{L} \left\langle\exp\Big[ 
\beta J^{(i_1,i_2),(j_1,j_2)}_{x,y} \sum_a S^a_{i_1}(x)S^a_{i_2}(x)
S^a_{j_1}(y)S^a_{j_2}(y)
\Big]\right\rangle
\Bigg] \nonumber \\
=& \mathrm{Tr}\exp\Bigg[
\sum_{(x, y)} \sum_{i_1<i_2}\sum_{j_1<j_2}
\ln\Bigg\{
1+\frac z L\Big[ 
\left\langle\exp\Big[ 
\beta J^{(i_1,i_2),(j_1,j_2)}_{x,y} \sum_a S^a_{i_1}(x)S^a_{i_2}(x)
S^a_{j_1}(y)S^a_{j_2}(y)
\Big]\right\rangle-1
\Big]
\Bigg\}
\Bigg] \nonumber\\
=& \mathrm{Tr}\exp\Bigg[\frac z L
\sum_{(x, y)} \sum_{i_1<i_2}\sum_{j_1<j_2}
\Big[ 
\left\langle\exp\Big[ 
\beta J^{(i_1,i_2),(j_1,j_2)}_{x,y} \sum_a S^a_{i_1}(x)S^a_{i_2}(x)
S^a_{j_1}(y)S^a_{j_2}(y)
\Big]\right\rangle-1
\Big],
\label{znbar}
\end{align}
where in the last line, we have neglected the terms of subleading order in $L$.

With the notation $\mathbb{J}\equiv J^{(i_1,i_2),(j_1,j_2)}_{x,y}$, we now write
\begin{align}
&  \left\langle  \exp\Big[ 
\beta \mathbb{J} \sum_a S^a_{i_1}(x)S^a_{i_2}(x) 
S^a_{j_1}(y)S^a_{j_2}(y)
\Big] \right\rangle\nonumber \\
=&\prod_a\left\langle \cosh\left(\beta\mathbb{J}\right)
\left\{1+  S^a_{i_1}(x)S^a_{i_2}(x) 
S^a_{j_1}(y)S^a_{j_2}(y) \tanh\left(\beta\mathbb{J}\right)
\right\}\right\rangle \nonumber \\
=& T_0 + T_2 \sum_{a<b}  S^a_{i_1}(x)S^a_{i_2}(x) S^b_{i_1}(x)S^b_{i_2}(x) 
S^a_{j_1}(y)S^a_{j_2}(y) S^b_{j_1}(y)S^b_{j_2}(y)\nonumber \\
+&T_4 \sum_{a<b<c<d}  S^a_{i_1}(x)S^a_{i_2}(x) S^b_{i_1}(x)S^b_{i_2}(x) 
 S^c_{i_1}(x)S^c_{i_2}(x) S^d_{i_1}(x)S^d_{i_2}(x) 
 S^a_{j_1}(y)S^a_{j_2}(y) S^b_{j_1}(y)S^b_{j_2}(y)
 S^c_{j_1}(y)S^c_{j_2}(y) S^d_{j_1}(y)S^d_{j_2}(y) \nonumber \\
 &+\cdots,
\end{align}
where
\begin{align}
    T_k\equiv \langle \cosh^n(\beta \mathbb{J})\tanh^k (\beta\mathbb{J})\rangle,
    \label{Tk}
\end{align}
and we have assumed that the distribution $P_z$ is symmetric under $J\to -J$ so that $T_k$ for odd
$k$ vanishes.
We note that we can write in the leading order in $L$ 
\begin{align}
    \sum_{(x, y)} \sum_{i_1<i_2}\sum_{j_1<j_2} S^a_{i_1}(x)S^a_{i_2}(x) S^b_{i_1}(x)S^b_{i_2}(x) 
S^a_{j_1}(y)S^a_{j_2}(y) S^b_{j_1}(y)S^b_{j_2}(y)
= \frac{1}{2} \Bigg[ \sum_x \sum_{i_1<i_2} S^a_{i_1}(x)S^a_{i_2}(x) S^b_{i_1}(x)S^b_{i_2}(x)\Bigg]^2
\end{align}
and
\begin{align}
   & \sum_{(x, y)} \sum_{i_1<i_2}\sum_{j_1<j_2}
  S^a_{i_1}(x)S^a_{i_2}(x) S^b_{i_1}(x)S^b_{i_2}(x) 
 S^c_{i_1}(x)S^c_{i_2}(x) S^d_{i_1}(x)S^d_{i_2}(x) 
 S^a_{j_1}(y)S^a_{j_2}(y) S^b_{j_1}(y)S^b_{j_2}(y)
 S^c_{j_1}(y)S^c_{j_2}(y) S^d_{j_1}(y)S^d_{j_2}(y) \nonumber \\  
 =&\frac 1 2 \left[ 
 \sum_x \sum_{i_1<i_2} 
 S^a_{i_1}(x)S^a_{i_2}(x) S^b_{i_1}(x)S^b_{i_2}(x) 
 S^c_{i_1}(x)S^c_{i_2}(x) S^d_{i_1}(x)S^d_{i_2}(x) 
 \right]^2.
\end{align}

We can now rewrite Eq.~(\ref{znbar}) as
\begin{align}
\langle Z^n\rangle_J=& \exp\left[\frac z L \frac{L^2}2 {M \choose 2}^2 (T_0-1)\right] \nonumber \\
\times&\mathrm{Tr}\Bigg\{ \exp\Bigg[ 
\frac{z}{2L}T_2 \sum_{a<b} 
\Big[ \sum_x \sum_{i_1<i_2} S^a_{i_1}(x)S^a_{i_2}(x) S^b_{i_1}(x)S^b_{i_2}(x)\Big]^2
\Bigg] \nonumber \\
&\times  
 \exp\Bigg[ 
\frac{z}{2L}T_4 \sum_{a<b<c<d} 
\Big[ \sum_x \sum_{i_1<i_2} S^a_{i_1}(x)S^a_{i_2}(x) S^b_{i_1}(x)S^b_{i_2}(x)
S^c_{i_1}(x)S^c_{i_2}(x) S^d_{i_1}(x)S^d_{i_2}(x)\Big]^2 +\cdots
\Bigg]
\Bigg\}.
\end{align}

We introduce the delta functions
enforcing
\begin{align}
q_{ab}=\frac 1 {LM^2} \sum^L_x \sum_{i_1<i_2}^M S^a_{i_1}(x)S^a_{i_2}(x)S^b_{i_1}(x)S^b_{i_2}(x) 
\end{align}
and
\begin{align}
    q_{abcd}=\frac 1 {LM^2} \sum^L_x \sum_{i_1<i_2}^M S^a_{i_1}(x)S^a_{i_2}(x)S^b_{i_1}(x)S^b_{i_2}(x)
    S^c_{i_1}(x)S^c_{i_2}(x)S^d_{i_1}(x)S^d_{i_2}(x),
\end{align}
etc.. We have
\begin{align}
\langle Z^n\rangle_J=e^{nL C}&\mathrm{Tr}\int\prod_{a <b}   dq_{ab} \prod_{a <b<c<d}   dq_{abcd}\cdots \; 
\exp \Big[ \frac {z L M^4} 2 T_2 \sum_{a< b}q^2_{ab} 
+ \frac {z L M^4} 2 T_4 \sum_{a< b<c<d}q^2_{abcd}+\cdots\Big] \nonumber \\
&\times\prod_{a<b}\delta\left(MLq_{ab}-\frac 1 M \sum^L_x \sum^M_{i<j} S^a_i (x)S^a_j (x) S^b_i(x) S^b_j(x)   \right) \nonumber \\
&\times\prod_{a<b<c<d}\delta\left(MLq_{abcd}-\frac 1 M \sum^L_x \sum^M_{i<j} 
S^a_i (x)S^a_j (x) S^b_i(x) S^b_j(x) 
S^c_i (x)S^c_j (x) S^d_i(x) S^d_j(x)\right) ,
\end{align}
where
\begin{align}
    C\equiv \frac z 2  {M \choose 2}^2 \lim_{n\to 0}\frac 1{n}(T_0-1)\to  \frac z 2  {M \choose 2}^2
    \langle \ln\cosh(\beta \mathbb{J})\rangle .
\end{align}

Using the integral representation of the delta function, we can write
\begin{align}
\langle Z^n\rangle_J=&e^{nLC}\int\prod_{a<b}   dq_{ab}   d\mu_{ab} \prod_{a<b<c<d}   dq_{abcd}d\mu_{abcd} \cdots \exp \Big[\frac {z L M^4} 2 T_2 \sum_{a< b}q^2_{ab}  -{M}L\sum_{a< b} \mu_{ab}q_{ab} \Big]\nonumber \\
\times & 
\exp\Big[\frac {z L M^4} 2 T_4 \sum_{a< b<c<d}q^2_{abcd}-ML \sum_{a<b<c<d}\mu_{abcd}q_{abcd} +\cdots
+L \ln \mathcal{L}(\underline{\mu})\Big]  \nonumber \\
=&e^{nLC}\int\prod_{a<b} dq_{ab}  d\mu_{ab} \prod_{a<b<c<d}   dq_{abcd}d\mu_{abcd} \cdots \;\exp[-LG(
\underline{q},\underline{\mu})], \label{zn}
\end{align}
where
\begin{equation}
G(
\underline{q},\underline{\mu})=
-\frac {z  M^4} 2 T_2 \sum_{a< b}q^2_{ab}  +{M}\sum_{a< b} \mu_{ab}q_{ab}
-\frac {z  M^4} 2 T_4 \sum_{a< b<c<d}q^2_{abcd}+M \sum_{a<b<c<d}\mu_{abcd}q_{abcd} +\cdots
- \ln \mathcal{L}(\underline{\mu})
\label{G}
\end{equation}
and 
\begin{align}
\label{L}
\mathcal{L}(\underline{\mu})=\underset{\{S_i^a\}}{\mathrm{Tr}} \exp\Big[   \frac 1 {M} \sum_{a< b} \mu_{ab}\sum^M_{i<j}
 S^a_i S^a_j  S^b_i S^b_j +\frac 1 M 
 \sum_{a< b<c<d} \mu_{abcd}\sum^M_{i<j}
 S^a_i S^a_j  S^b_i S^b_jS^c_i S^c_j  S^d_i S^d_j +\cdots
 \Big].
\end{align}

The integral is dominated by the saddle points which are determined by
\begin{align}
\mu_{ab}=z M^3 T_2 q_{ab},~~~~~
\mu_{abcd}=z M^3 T_4 q_{abcd}\label{lq}
\end{align}
and 
\begin{align}
q_{ab}=\frac 1 {M^2} \left\langle  \sum^M_{i<j} S^a_i S^a_j  S^b_i S^b_j \right\rangle_\mathcal{L} ,
~~~~~
q_{abcd}=\frac 1 {M^2} \left\langle  \sum^M_{i<j} S^a_i S^a_j  S^b_i S^b_j
S^c_i S^c_j  S^d_i S^d_j\right\rangle_\mathcal{L}
\end{align}
where $\langle\cdots\rangle_\mathcal{L}$ is evaluated with respect to $\mathcal{L}$ in Eq.~(\ref{L}).
Then the free energy $F$ is given by
\begin{align}
\frac{\beta F}{L}&=-\frac 1 L \lim_{n\to 0}\frac 1 n \ln \langle Z^n\rangle_J 
=-C +\lim_{n\to 0}\frac 1 n G(\underline{q},\underline{\mu}).
\label{free1}
\end{align}

When we expand $\mathcal{L}$ to cubic order in powers of $\underline{\mu}$ and use Eqs.~(\ref{lq}), we have
\begin{align}
\frac{\beta F}{L} &\simeq   -C  -M\ln 2  \\
&+\lim_{n\to 0}\frac 1 n \Big[ \tau_2
\sum_{a< b} q^2_{ab}+ \tau_4 \sum_{a<b<c<d} q^2_{abcd}
-w_1 \sum_{(a,b,c)}q_{ab}q_{bc}q_{ca} 
- w_2 \sum_{a\neq b}q^3_{ab} \nonumber \\
&~~~~~~~~~   -w_3 \sum_{(a,b,c,d)} q_{ab}q_{cd}q_{abcd} 
-w_4\sum_{(a,b,c,d,e)} q_{ab}q_{acde}q_{bcde}
-w_5 \sum_{(a,b,c,d)} q^3_{abcd} \nonumber \\
&~~~~~~~~~ -w_6 \sum_{(a,b,c,d,e,f)}q_{abcd}q_{cdef}q_{efab}
\Big], 
\end{align}
where
\begin{align}
&    \tau_2 = \frac{z}{2}M^4 T_2 \left\{ 1 -z {M\choose 2} T_2\right\} \\
& \tau_4 = \frac{z}{2}M^4 T_4 \left\{ 1 -z {M\choose 2} T_4\right\} .
\end{align}
From Eq.~(\ref{Tk}), we can easily see that $T_2>T_4$. Therefore, at temperature $\tau_2=0$, we have $\tau_4>0$.
Therefore, $T_c$ in this model is determined by $\tau_2=0$ or 
\begin{align}
   z {M\choose 2}  \left\langle \tanh^2 (\beta_c\mathbb{J})\right\rangle =1.
    \label{Tc_VB}
\end{align}
When $P_z$ is given by the Gaussian distribution of zero mean
and variance $1/M^3$, we can easily solve the above equation. 
For $M=4$ and $z=6$, we have $T_c=1/\beta_c=0.729$.

\section{Exact expression for the spin-glass susceptibility in the $M=4$, $p=4$ model}
\label{appendix:susceptibility}

In this appendix, we derive the exact expression for the spin-glass susceptibility for the balanced $M=4$, $p=4$ model using the replicon formalism \cite{AJBray_1979}. As discussed in the main text, the spin-glass susceptibility is related to the replicon propagator and is given by
\begin{align}
\chi_{\mathrm{SG}}(k)=G_1-2G_2+G_3,
\end{align}
where
\begin{align}
G_1 &= \langle q_{ab}(i)\,q_{ab}(j)\rangle, \\
G_2 &= \langle q_{ab}(i)\,q_{ac}(j)\rangle, \\
G_3 &= \langle q_{ab}(i)\,q_{cd}(j)\rangle.
\end{align}

For the $M=4$ model, the local overlap field between replicas $a$ and $b$ at site $i$ is defined as
\begin{align}
q_{ab}(i)=\frac{1}{M^2}
\Big[
S_i^aT_i^aS_i^bT_i^b
+S_i^aU_i^aS_i^bU_i^b
+S_i^aV_i^aS_i^bV_i^b
+T_i^aU_i^aT_i^bU_i^b
+T_i^aV_i^aT_i^bV_i^b
+U_i^aV_i^aU_i^bV_i^b
\Big].
\end{align}
Thus, each rung contains $\binom{4}{2}=6$ independent pair variables,
\[
(ST,\;SU,\;SV,\;TU,\;TV,\;UV),
\]
and therefore each pair of sites contributes a total of $6\times 6=36$ terms to the susceptibility.

We first evaluate $G_1$, where the same replica pair appears at both sites:
\begin{align}
G_1=\langle q_{ab}(i)\,q_{ab}(j)\rangle.
\end{align}
Substituting the overlap field and expanding the product, we obtain all possible combinations of the six local pair operators at sites $i$ and $j$. After thermal averaging and using replica factorization, this becomes
\begin{align}
M^4G_1=
\sum_{\text{36 terms}}
\langle \mathcal{O}_i\mathcal{O}_j\rangle^2,
\label{eq:G1_appendix}
\end{align}
where $\mathcal{O}_i$ denotes one of the six local pair operators.

Next, for $G_2$, one replica index is common while the other differs:
\begin{align}
G_2=\langle q_{ab}(i)\,q_{ac}(j)\rangle.
\end{align}
In this case, the average factorizes into one two-point correlator and two one-point averages due to replica independence. Summing over all possible pair combinations gives
\begin{align}
M^4G_2=
\sum_{\text{36 terms}}
\langle \mathcal{O}_i\mathcal{O}_j\rangle
\langle \mathcal{O}_i\rangle
\langle \mathcal{O}_j\rangle.
\label{eq:G2_appendix}
\end{align}

For $G_3$, all replica indices are different:
\begin{align}
G_3=\langle q_{ab}(i)\,q_{cd}(j)\rangle.
\end{align}
Here the thermal average factorizes completely, leading to
\begin{align}
M^4G_3=
\sum_{\text{36 terms}}
\langle \mathcal{O}_i\rangle^2
\langle \mathcal{O}_j\rangle^2.
\label{eq:G3_appendix}
\end{align}

Combining Eqs.~\eqref{eq:G1_appendix}--\eqref{eq:G3_appendix}, the zero-momentum spin-glass susceptibility becomes
\begin{align}
\chi_{\mathrm{SG}} \equiv \chi_{\mathrm{SG}}(0)
=
\frac{1}{LM^4}
\sum_{ij}
\Bigg[
\sum_{\text{36 pairs}}
\Big(
\langle \mathcal{O}_i\mathcal{O}_j\rangle
-
\langle \mathcal{O}_i\rangle
\langle \mathcal{O}_j\rangle
\Big)^2
\Bigg]_{\mathrm{av}}.
\end{align}

This is the exact expression for the spin-glass susceptibility used in the main text for the $M=4$, $p=4$ model. The summation over all $36$ pair combinations arises naturally from the six possible local pair variables on each rung.

\twocolumngrid
\bibliography{references}
\end{document}